\documentclass[aps,prx,twocolumn,superscriptaddress,longbibliography,floatfix]{revtex4-2}

\usepackage[T1]{fontenc}
\usepackage[utf8]{inputenc}
\usepackage{amsmath,amssymb,amsfonts}
\usepackage{bm}
\usepackage{physics}
\usepackage{graphicx}
\usepackage{xcolor}
\usepackage{hyperref}
\usepackage{siunitx}
\hypersetup{
  colorlinks=true,
  linkcolor=blue,
  citecolor=blue,
  urlcolor=blue
}

\newcommand{\Diff}{\mathrm{d}}

\newcommand{\e}{\mathrm{e}}
\def\bea{\begin{eqnarray}}

\newcommand{\defas}{\ensuremath{:=}}

\graphicspath{ {./images/} }

\begin{document}

\title{Rotational Quantum Tunneling of a Magnetic Dipole in a Superconducting Trap}

\author{Francis J. Headley}
\thanks{These authors contributed equally}
\affiliation{Institut für Theoretische Physik, Eberhard-Karls-Universität Tübingen, 72076 Tübingen, Germany}
\email{francis.headley@uni-tuebingen.de}

\author{Fabian Müller}
\thanks{These authors contributed equally}
\affiliation{Institut für Theoretische Physik, Eberhard-Karls-Universität Tübingen, 72076 Tübingen, Germany}
\affiliation{Department of Condensed Matter Physics, Charles University, 121 16 Prague 2, Czech Republic}
\email{fabian.muller@matfyz.cuni.cz}

\author{Emre Köse}
\thanks{These authors contributed equally}
\affiliation{Institut für Theoretische Physik, Eberhard-Karls-Universität Tübingen, 72076 Tübingen, Germany}
\affiliation{Instituto de Física Corpuscular (IFIC), CSIC–Universitat de València, and Departament de Física Teòrica, UV, Parc Científic UV, c/ Catedrático José Beltrán, 2, E-46980 Paterna (València), Spain}
\email{emrekose@ific.uv.es}

\author{Tim Fuchs}
\affiliation{School of Physics and Astronomy, University of Southampton, SO17 1BJ, Southampton, UK}

\author{Hendrik Ulbricht}
\affiliation{School of Physics and Astronomy, University of Southampton, SO17 1BJ, Southampton, UK}

\author{Daniel Braun}
\affiliation{Institut für Theoretische Physik, Eberhard-Karls-Universität Tübingen, 72076 Tübingen, Germany}
\email{daniel.braun@uni-tuebingen.de}

\date{\today}
\begin{abstract}
  We study the quantum dynamics of the rotational degree of freedom of a nano-magnet trapped in a superconducting trap. The nano-magnet is modeled as a magnetic dipole with magnetization pinned to the easy axis of the particle.  The magnetic trap then leads to a potential barrier that hinders free rotation of the particle, but through which it can tunnel. We identified rest-gas scattering as the most important decoherence mechanism at low temperatures. A shape of the particle sufficiently close to perfect rotational symmetry about the rotational axis can protect the rotational tunneling against this decoherence mechanism,  and we identify experimentally feasible parameter regimes where rotational tunneling should be observable.
\end{abstract}
\maketitle

\section{Introduction}

The phenomenon of quantum tunneling, a hallmark of quantum mechanics,
enables particles to traverse energy barriers that would otherwise be
classically forbidden.
Rotational tunneling was intensively investigated in molecular physics
starting in the 1970s \cite{KollmarAlefeld1976,Hueller1977,HuellerPress1981,Stevens1983,Haeusler85,hausler_zur_1989}.  There, methyl groups attached to
macromolecules can rotate around the bond that attaches it to the
macromolecule, but experience a periodic hindering potential, due to
interaction with the rest of the molecule, through which they can tunnel.
The molecules are embedded in a molecular solid, creating a thermal
phonon bath and an open quantum system for the methyl group.  
Neutron scattering experiments show that 
tunneling splittings can be observed up to temperatures of several
10\,K, much larger than the 
tunneling splitting.  Decoherence only sets in at temperature scales comparable
to the librational frequency of the methyl group in the hindering potential.  Based on
microscopic models of the open quantum system, it was realized that
the coherence of the tunneling was protected by what later in quantum
information science would be called a decoherence-free subspace (DFS) \cite{Braun1993,Braun1994}: The
coupling to the environment has the same symmetry as the hindering potential and this prevents it from
distinguishing states in the DFS. Transitions within the DFS are
symmetry-forbidden, and decoherence only sets in due to transitions driven by the heatbath to the next
librational energy eigenstates, corresponding to much higher temperatures.

In recent years, rotational quantum coherence has attracted renewed interest
in the context of levitated nanoparticles \cite{Zhong2016,Stickler2018,martinetz_quantum_2020,gonzalez-ballestero_levitodynamics_2021,vinante_levitated_2022,pedernales_decoherence-free_2020,carlesso_perturbative_2021,stickler_quantum_2021,Glikin2025}.  Notably, magnetic
nanoparticles trapped in superconducting traps can
have preferred orientations of the particle between which it is expected to 
tunnel in a suitable parameter regime. This links to an
older research line on macroscopic quantum tunneling of the
magnetization direction in small magnetic particles, where the
particle is kept fixed but the magnetization can rotate \cite{barbara_quantum_1997,wernsdorfer_exchange-biased_2002,wernsdorfer_classical_2006}.  The
potential height can be controlled by magnetic fields, and in the limit
of single-domain magnetization the potential energy part of such a system is described by the
Stoner-Wolfarth model \cite{stoner_mechanism_1948}.  Its energy barriers can be calculated exactly
and play an important role technologically in integrated magnetic
memories (MRAM), where thermally activated switching must be
suppressed \cite{braun_exact_2004}. 

Substantial work has been put in understanding the decoherence of
rotational degrees of freedom of suspended nanoparticles
\cite{carlesso_perturbative_2021,Zhong2016,stickler_rotational_2018,stickler_spatio-orientational_2016}. Depending on the couplings to the
environment, one expects that the
DFS protection of coherences in rotational degrees of freedom still
works at least partially. Suspended magnetic nanoparticles brought
into quantum superpositions of their rotational degree of freedom by
quantum tunneling might then
offer a unique platform to probe fundamental physics, including precision measurements and searches for new physics such as dark matter or quantum gravitational effects 
\cite{carney_tabletop_2019,bose_spin_2017,headley_quantum_2025}.

\begin{figure}[t!]
  \centering
  \includegraphics[width=0.9\linewidth]{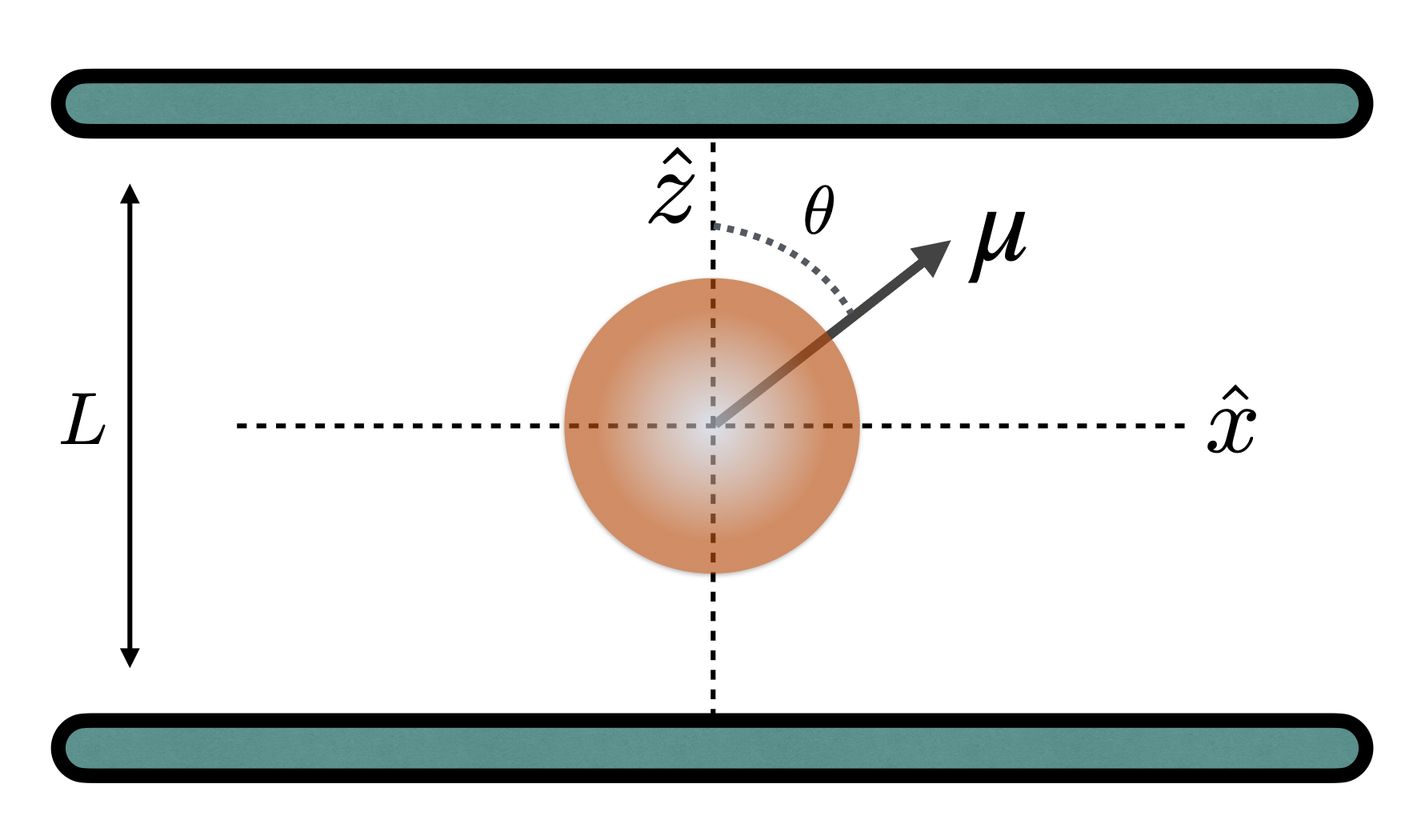}
  \caption{Sketch  
  of the setup. A nanomagnet, ideally of spherical shape, with magnetic moment $\bm{\mu}$ is trapped between two superconducting plates.} 
  \label{fig:diag}
\end{figure}

In this context, we investigate in the present work the quantum
tunneling dynamics of a magnetic dipole confined between two infinite
parallel superconducting plates, a setup that yields an analytic
periodic potential with $C_2$-symmetry, ideal for exploring tunneling phenomena. The trapping potential,
derived in \cite{headley_magnetic_2025} using the method of image dipoles, results in a double-well structure with minima corresponding to the dipole aligned parallel to the plates, which can further be modulated by a perpendicular magnetic field, similar to the Stoner-Wohlfarth model.

\section{Theoretical Model}

We consider a magnetic dipole levitated between two infinite parallel
superconducting plates located at $z = L/2$ and $z = -L/2$, where the
$z$-axis is defined as perpendicular to superconducting plates. We restrict
ourselves to rotations of the dipole in a fixed plane ($x$-$z$--plane), which can be
achieved with a second, narrow pair of parallel superconducting plates
perpendicular to the first one. The dipole moment can then be
parameterized with a single angle $\theta$ measured from the $z$-axis,
with $x,z$ components given by $\boldsymbol{\mu} = \mu (\sin \theta, \cos \theta)$.
The dipole is situated at $\mathbf{r}_0 = (0, z_0=0)$, taken as fixed in
the following. The plates are modeled as perfect diamagnets in the
Meissner state, enforcing the boundary condition that the normal
component of the magnetic field vanishes at the plate surfaces
($\mathbf{n} \cdot \mathbf{B} = 0$), see Fig.~\ref{fig:diag}. 

The potential experienced by the dipole results from
its interaction with the magnetic field induced by surface currents on
the superconducting plates, which 
can be computed using the method of image
dipoles. The total magnetic scalar potential $\Phi_{\text{tot}}$ is obtained by
summing the contributions from the physical dipole and an infinite
series of image dipoles, which satisfy the boundary conditions at both
plates. The potential energy of the dipole is given by $U(z, \theta) =
-\frac{1}{2} \boldsymbol{\mu} \cdot \mathbf{B}_{\text{I}}$, where
$\mathbf{B}_{\text{I}} = -\mu_0 \nabla \Phi_{\text{I}}$ is the
magnetic field due to the image dipoles, evaluated at the dipole's
position. For a dipole centered symmetrically between 
plates at $z_0 = 0$, the rotational potential reduces,
up to an arbitrary constant to \cite{headley_magnetic_2025}
\begin{equation}
U(\theta) = \frac{V_0}{2}\left( 1 +\cos (2 \theta)
\right),\,\,V_0=\frac{\mu_0 \mu^2\zeta(3)}{8 L^3 \pi}, \label{eq:ogpot}
\end{equation}
where $\zeta(n)$ is the Riemann zeta function for $n\in\mathbb{Z}$. This potential exhibits two  minima at $\theta = \pm \pi/2$, corresponding to the dipole aligned parallel to the $x$-axis, between which it can tunnel if the potential barrier between them is not too high.  
We assume that an  additional magnetic field with components $B_x,B_z$ in the $x,z$-direction can be applied. These are total fields, including the Meissner effect from the superconducting plates. 

The full Hamiltonian for the rotational degree of freedom then reads
\begin{equation}
H = -\frac{\hbar^2}{2I} \partial^2_\theta + \frac{V_0}{2} \left( 1 + \cos (2 \theta) \right) -\mu B_z\cos\theta -\mu B_x \sin \theta\,, \label{ham} 
\end{equation}
where $I=\frac{2}{5}m R^2$ is the moment of inertia of the spherical {nanomagnet of mass $m$ and radius $R$}. 
Without the kinetic energy, the model reduces to the Stoner-Wolfarth
model of single-domain magnetic particles in a magnetic field, with
its characteristic stability astroid and hysteretic switching behavior
\cite{stoner_mechanism_1948}: 
The magnetic field $B_x$ parallel to the superconducting plates
creates an asymmetry between the height
of the local minima of the two potential wells, and for sufficiently
strong $B_x$ the higher local minimum disappears, resulting, for $B_z=0$,
in abrupt and hysteretic alignment of the magnetization with the applied field when
it was previously antiparallel to the field.  The magnetic
field $B_z$ perpendicular to the
superconducting plates reduces the height of the potential barrier
in one rotational direction and increases it in the other. At $B_x=0$,
the minima remain degenerate in energy and magnetization tends to tilt
more and more in the direction of $B_z$.  For sufficiently strong
$B_z$ the two minima coalesce and only one stable magnetization
remains. 

With $V_0$, $B_x$, and $B_z$ one has large control over the potential barrier and the tunneling motion.  In the following,  we first consider the undamped dynamics and derive from it already bounds on the parameters that allow tunneling motion, such as the combinations of $v_0,h_z$ that assure that the energy eigenvalues $E_0,E_1$ remain both below the potential-barrier height. The bounds will be further refined when we include coupling to the environment and decoherence, explored in Sec.~\ref{sec:dec}.

\subsection{External-field-free case}\label{sec.FFC}
\begin{figure}[t!]
  \centering
  \includegraphics[width=0.9\linewidth]{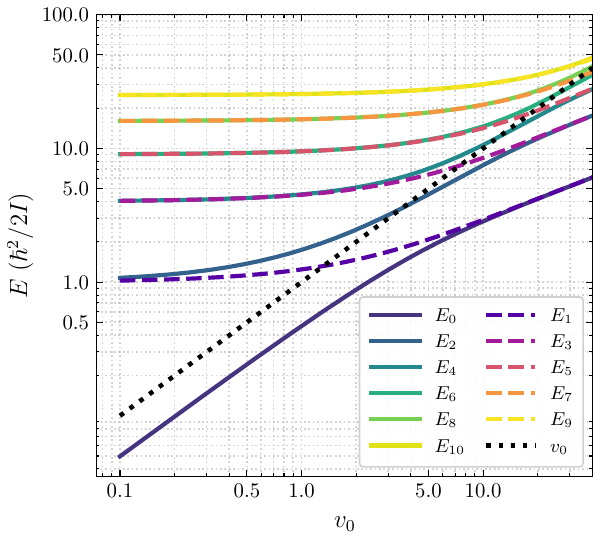}
  \caption{Energy spectrum (in units of $E_\text{k}$) of the external field-free rotor
    as function of the parameter $v_0$. For $v_0\to 0$, one finds doubly degenerate
    angular momentum eigenstates, corresponding to clockwise or
    anti-clock wise rotation. 
    The lowest one
    with angular momentum 0 is non-degenerate.  For $v_0\gtrsim 1.3$, the two lowest energy levels
    remain under the potential barrier 
    (straight black dotted line) and correspond to a
    tunneling-split ground state. For $v_0\gg 1$, tunneling
    is exponentially suppressed, leading to doubly degenerate states
    localized in one or the other well. }

  \label{fig:ews}
\end{figure}

In the case $B_x=B_z=0$, the stationary Schr\"odinger equation for an
energy eigenstate $\psi$ in angle representation,
$\psi(\theta)=\braket{\theta}{\psi}$ and eigenenergy $E$ reads
\begin{equation}
  \label{eq:SE}
  \partial_\theta^2\psi(\theta)+\left(\frac{2I}{\hbar^2}(E-\frac{V_0}{2})-\frac{I
    V_0}{\hbar^2 }\cos(2\theta)\right)\psi(\theta)=0\,.
\end{equation}
With the identifications
\begin{equation}
  \label{eq:idents}
a=\frac{2I}{\hbar^2}(E-\frac{V_0}{2}), \,\,\,q=\frac{I V_0}{2\hbar^2}\,,  
\end{equation}
it maps to the Mathieu equation
\begin{equation}
  \label{eq:Mathieu}
   \partial_\theta^2\psi(\theta)+\left(a-2q\cos(2\theta)\right)\psi(\theta)=0\,,
 \end{equation}
 whose solutions are exactly known. They lead to energy eigenstates
 and eigenvalues given by
 \begin{eqnarray}
   \label{eq:psisM}
   \psi_{2n}(\theta)&=& \text{Ce}_n(a_n(q),q,\theta),
                           \,\,\,E_{2n}=\frac{\hbar^2
                           }{2I}a_n(q)+\frac{V_0}{2}\\
   \psi_{2n-1}(\theta)&=& \text{Se}_n(b_n(q),q,\theta),
                           \,\,\,E_{2n-1}=\frac{\hbar^2
                           }{2I}b_n(q)+\frac{V_0}{2}\,,\nonumber
 \end{eqnarray}
where the $\text{Ce}_n$ and $\text{Se}_n$ are the even and odd Mathieu functions of the
first kind, respectively,  and $a_n(q)$, $b_n(q)$
are the corresponding characteristic numbers with $n=0,1,2,3,\ldots$,
for the even case and $n=1,2,3,\ldots$ in the odd case.   
 In the following we use as energy scale the quantum of kinetic energy $E_\text{k}\defas\hbar^2/(2I)$
and express all energies in this unit. Fig.~\ref{fig:ews} shows the
energy eigenvalues as function of $v_0\defas V_0/E_\text{k}$. For
$v_0\to 0$, the rotor becomes unhindered and we get back the energy
angular momentum eigenstates $\psi_n(\theta)=e^{i n
  \theta}/\sqrt{2\pi}$ with eigenvalues $e_n\defas
E_n/E_\text{k}=n^2$, $n=0,\pm 1,\pm 2,\ldots$, all twice degenerate with the exception of
$n=0$.  For $v_0\to\infty$, two infinitely deep potential wells arise,
tunneling is exponentially suppressed, and the lowest energy levels
can be obtained from their harmonic approximations around the minima,
leading to double-degeneracy of all levels.  

\begin{figure}[t!]
  \centering
  \includegraphics[width=0.85\linewidth]{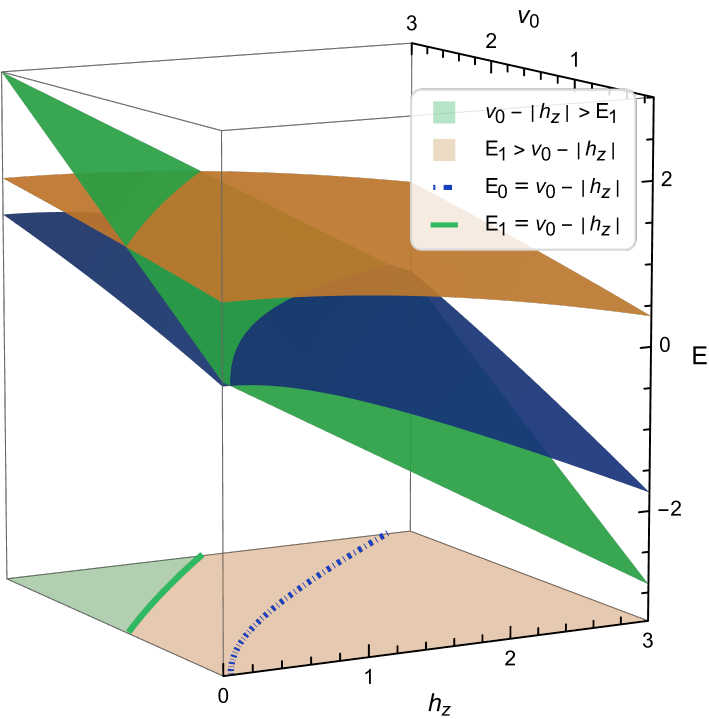}
  \caption{ Energy eigenvalues $E_0$ (blue surface) and $E_1$ (orange surface) of the two lowest states, together with the approximate lower potential-barrier height $v_0-|h_z|$ (green surface), plotted as functions of the dimensionless barrier-height parameter $v_0$ and the perpendicular-field parameter $h_z$ (with $h_x = 0$). When the first excited level $E_1$ rises above the green surface, the particle can classically surmount the lower barrier and the system exits the tunneling regime. The intersection curve of the $E_1$ and barrier surfaces therefore delineates the boundary of the parameter region in which coherent tunneling between the two orientational wells is possible. For $h_z = 0$, this crossing occurs near $v_0\approx1.3$. A figure showing the regions where $v_0-|h_z|>E_1$ is given in the Appendix (c.f. Fig.~\ref{fig:crossing_analysis})}
  \label{fig:3d_energy}
\end{figure}

\begin{figure*}[t!]
  \centering
  \includegraphics[width=0.99\linewidth]{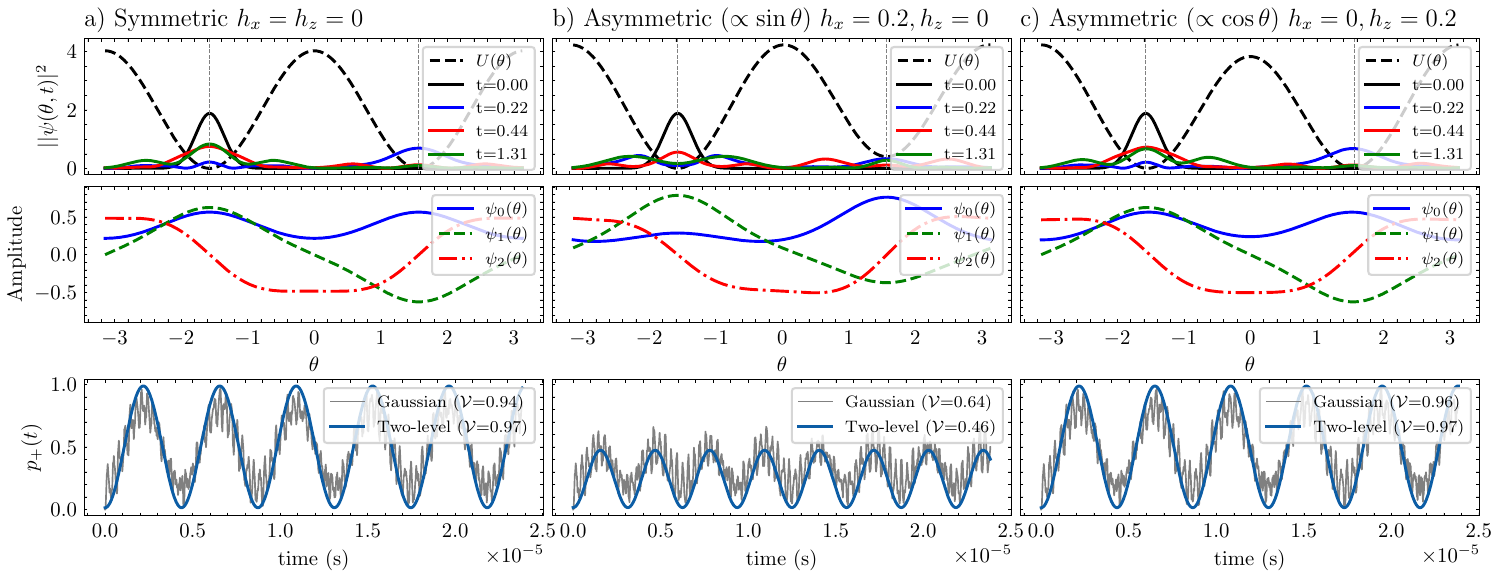}
  \caption{Figs.~(a), (b) and (c) describes the potentials without any external field, and with external fields parallel and perpendicular to superconducting plates, respectively. First row of figures: 
  Dimensionless potential as a function of $\theta$ for each scenario (dashed black) and the evolution of the state chosen to be an approximate Gaussian wave-packet for different times $t=\{0,1.19,2.38,4.77\}$ in units of $10^{-5}$ seconds (solid color). 
  Second row of figures: 
  Wave functions for the ground state, the first exited state and second excited states. The last row shows the evolution of the $p_+(t)$ 
  unitary dynamics for different potential cases for an initial state chosen to be a Gaussian wave-packet (gray) or two-level superposition state which localizes 
  the particle in left potential well. The 
  other  
   parameters are given in the Table \ref{tab:physical_correspondence}.}
  \label{fig:3}
\end{figure*}

\subsection{Case of additional applied magnetic fields}
In dimensionless quantities, Eq.~\eqref{ham} reads
\begin{equation}
H/E_\text{k}= - \partial^2_\theta + \frac{v_0}{2} \left( 1 + \cos (2 \theta) \right) -h_z\cos\theta - h_x \sin \theta\,, \label{ham2} 
\end{equation}
with $h_i=\mu B_i/E_\text{k}$, $i=x,z$. The energy eigenstates $\ket{\psi_m}$ and eigenvalues $E_m$ ($m=0,1,2,\ldots$) and hence the undamped dynamics can be computed numerically by exact diagonlization.  For this, we represent the Hamiltonian in the basis $ \psi_n(\theta)$ for $n\in -N,-N+1,\ldots,N$ up to some cut-off $N$, with  $N = 200$ for unitary evolution and $N=20$ for the simulation of the master equation in the dissipative case. The two lowest energy eigenstates as function of $v_0$ and $h_z$ for $h_x=0$ are shown in Fig.~\ref{fig:3d_energy}.

For studying the dynamics of the tunneling, we consider two different initial states:  an approximately Gaussian wavepacket and a localized superposition of the two lowest energy-eigenstates split by the tunneling splitting. \\
\textit{Approximate Gaussian wavepacket.}
A wavepacket centered at the minimum of the potential at $\theta = -\pi/2$, can be constructed as 
\begin{equation}
\Psi_0(\theta) = \left\{
\begin{array}{cc}
     \mathcal{N} \exp\left(-\frac{(\theta + \pi/2)^2}{2\sigma^2}\right) & \text{ for} -\pi\le\theta\le 0, \\
    0 & \text{else.} 
\end{array}\right.
\end{equation}
The normalization constant $\mathcal{N}$ depends on the width $\sigma$ and assures $\int_{-\pi}^{\pi} |\Psi_0(\theta)|^2 d\theta = 1$. We choose $\sigma=0.3$, and the probability to be in the left well, $p_-(t)\defas
\int_{-\pi}^0|\Psi(\theta,t)|^2d\theta$, satisfies $p_-(0)=1$. Here, $\Psi(\theta,t)$ is the time-evolved state of the rotor in the absence of decoherence in $\theta$ representation with initial condition $\Psi(0,t)=\Psi_0$. \\
\textit{Two-level initial state.}
Given the structure of the two lowest energy eigenstates, also a  normalized superposition of the ground ($|\psi_0\rangle$) and first excited ($|\psi_1\rangle$) eigenstates of the system Hamiltonian,
\begin{equation}
|\Psi_0\rangle = \frac{|\psi_0\rangle + |\psi_1\rangle}{\sqrt{2}},
\end{equation}
approximates a state localized around $\theta =-\pi/2$.

\subsection{Generic Markovian Model} 
Magnetically levitated systems can sustain high quality factors \cite{fuchs_magnetic_2023, timberlake_acceleration_2019, vinante_ultralow_2020} and strong isolation from environmental disturbances \cite{gonzalez-ballestero_levitodynamics_2021,Mueller2025}, which is essential for stable platforms for precision measurements \cite{ahrens_levitated_2025}. Couplings to the environment can suppress the coherences between different energy eigenstates. If the coherence between the lowest tunneling-split eigenstates is destroyed, the visibility of tunneling oscillations is suppressed down to the extremely low temperatures corresponding to the energy splitting.  However, depending on the symmetry of the coupling to the environment and the symmetries of the energy eigenstates, transitions between the lowest tunneling-split eigenstates can be forbidden to first order perturbation theory.  The quantum coherence between  these states is then protected by a decoherence-free subspace, and decoherence of the tunneling motion kicks in only at much higher  temperatures when transitions to higher  energy eigenstates are excited. We discuss the symmetry aspects in detail in Sec.~\ref{sec.sym}.

Here we briefly summarize the derivation of the Markovian master equation; a
more detailed derivation following \cite{Breuer06} is given
in Appendix~\ref{app:master}. In the interaction picture 
with respect to the system-bath Hamiltonian $H_{\rm sys}+H_{\rm B}$, the interaction Hamiltonian
reads
\begin{equation}
  H_{\rm int}(t)
  = \sum_\alpha S^{(\alpha)}(t)\otimes B^{(\alpha)}(t),
\end{equation}
where the $S^{(\alpha)}$ are the ``system coupling agents'', i.e.~the
system operators in the interaction Hamiltonian and $B^{(\alpha)}$ are bath operators in the interaction picture \cite{Braun01,Strunz02}.
Under the standard Born and Markov approximations, and after performing
the secular approximation in the energy eigenbasis of $H_{\rm sys}$, one
obtains a Gorini--Kossakowski--Lindblad--Sudarshan (GKLS) equation for
the reduced density matrix,
\begin{equation}
\label{eq:ME-main}
\frac{\partial\rho}{\partial t}
= -\frac{i}{\hbar}[H_{\rm sys}+H_{\rm LS},\rho]
+ \sum_\alpha \sum_{\omega}
\gamma_\alpha(\omega)\,
\mathcal{D}\!\left[S^{(\alpha)}(\omega)\right]\rho,
\end{equation}
where $\mathcal{D}[L]\rho = L\rho L^\dagger - \tfrac{1}{2}
\{L^\dagger L,\rho\}$, $H_{\rm LS}$ is a Lamb-shift Hamiltonian, and
$S^{(\alpha)}(\omega)$ are the spectral components of the system
operators in the energy basis. Writing the eigenvalue equation
$
  H_{\rm sys}\ket{\psi_m} = E_m\ket{\psi_m},
$
with Bohr frequencies $\omega_{mn}=(E_m-E_n)/\hbar$, we define
\begin{equation}
  S^{(\alpha)}(\omega)
  =
  \sum_{E_m-E_n=\hbar\omega}
  \ket{\psi_m}\bra{\psi_m}
  S^{(\alpha)}
  \ket{\psi_n}\bra{\psi_n}. \label{eq:jumps_main_energybasis}
\end{equation}
The corresponding rates
$\gamma_\alpha(\omega)=2\,\mathrm{Re}\,\Gamma_\alpha(\omega)$ are
obtained from the one-sided Fourier transforms of the bath correlation
functions,
\begin{equation}
\Gamma_\alpha(\omega) = \frac{1}{\hbar^2}\int_0^\infty \Diff\tau\,
\e^{i\omega\tau}\,
\langle B^{(\alpha)}(\tau)B^{(\alpha)}(0)\rangle.
\end{equation}
For a bosonic bath at temperature $T_\alpha$ with spectral density
$J_\alpha(\omega)$ one finds
\begin{equation}
  \gamma_\alpha(\omega)
  = \frac{2\pi J_\alpha(|\omega|)}{\hbar^2}
    \begin{cases}
      n_\alpha(|\omega|)+1, & \omega>0,\\[2pt]
      n_\alpha(|\omega|),   & \omega<0,
    \end{cases}
\end{equation}
where $n_\alpha(\omega) = [\exp(\hbar\omega/k_B T_\alpha)-1]^{-1}$ is
the mean Bose--Einstein occupation number. The $\omega\neq 0$ terms describe
transitions between energy eigenstates 
of different energy, while the $\omega=0$ component
$S^{(\alpha)}(0)$ 
couples energy eigenstates that are degenerate in energy. 
In our numerical implementation we approximate each $J_\alpha(\omega)$
as effectively constant over the narrow band of Bohr frequencies relevant for the tunneling dynamics and absorb it into effective decay rates $\Gamma_\alpha$ with different coupling operators $\alpha \in \{ \cos\theta, \sin\theta,\cos2\theta,\sin2\theta,L_\theta\}$.

\section{Results and parameter analysis}

\subsection{Decoherence-free case}
We  project $\Psi(\theta,t)$ onto the energy eigenbasis to determine the
expansion coefficients. The time evolution of the wavepacket is then
calculated to determine the probability $p_+(t)=1-p_-(t)$ of the dipole tunneling to the
opposite well centered at $\theta = \pi/2$.  
To what extend tunneling takes places can be judged based on the
``visibility'' of the tunneling motion, defined as 
\begin{equation}
  \label{eq:V}
  \mathcal{V}={\text{max}(p_+(t))-\text{min}(p_+(t))}
\end{equation}
where the maximization and minimization are over
a time interval {starting at the time of the first maximum of $p_+(t)$} and larger than the inverse tunneling frequency at
the chosen parameter values. If no maximum of $p_+(t)$ exists, e.g.~for a pure thermal relaxation towards the equilibrium value $p_+=1/2$, we set $\mathcal{V}=0$.

In Fig.~\ref{fig:3} we compare the coherent tunneling dynamics for three different potential landscapes: the symmetric double well with no applied field, an asymmetric case generated by a field parallel to the plates ($h_x\neq 0$) and another asymmetric case produced by a perpendicular field ($h_z\neq 0$). In the first row, the dashed curve gives the potential and the colored curves show the time evolution of the initially localized wave packet. In the symmetric case, the wave packet moves back and forth between the two wells in a nearly balanced way, which is the clearest signature of coherent tunneling. When 
$h_x$ is applied, the left-right symmetry is broken, so one well becomes energetically preferred and the oscillations become less balanced. By contrast, a finite $h_z$ mainly distorts the barrier shape while keeping the wells nearly comparable in energy, therefore strong coherent oscillation is preserved.

The second and third rows of Fig.~\ref{fig:3} clarify the microscopic origin of this behavior. The plotted eigenfunctions show how the lowest states are distributed across the two wells, while the bottom row shows the probability $p_+(t)$ of finding the rotor in the right well. In all three cases, both the Gaussian initial state and the two-level superposition produce oscillations, with visibilities close to 1 for the Figs.~\ref{fig:3}(a) and \ref{fig:3}(c) and highly reduced for Fig.~\ref{fig:3} (b). This demonstrates that the tunneling dynamics is dominated by the lowest doublet and is therefore well described as coherent motion between two 
minima of the potential well. The reduction in visibility for the   asymmetric case with $h_x\ne 0$ reflects the field-induced bias between the wells, whereas the symmetric and the $h_z\ne 0$ asymmetric cases give almost ideal tunneling visibility.

\begin{figure}[t!]
  \centering
  \includegraphics[width=0.515\linewidth]{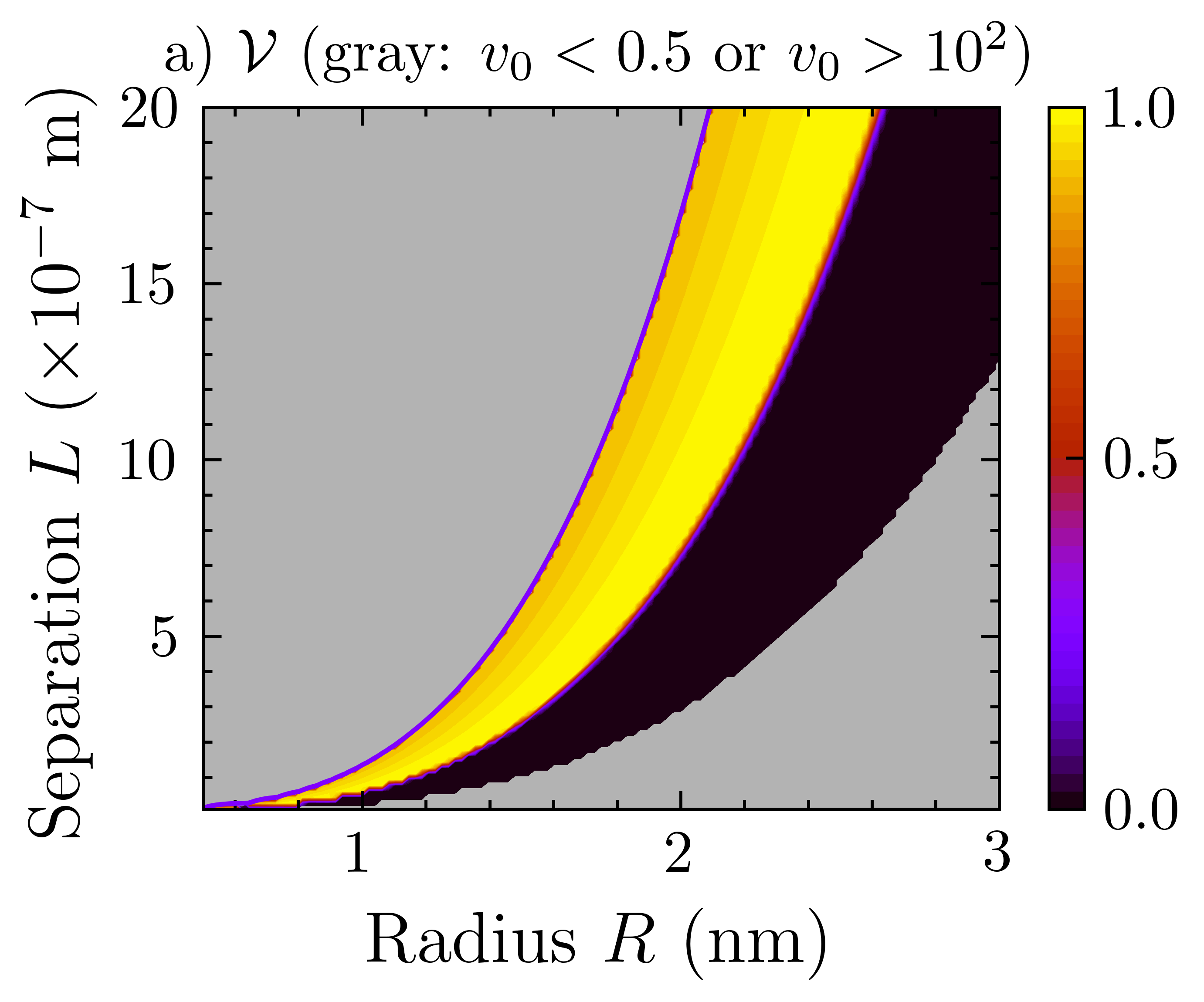}
  \includegraphics[width=0.482\linewidth]{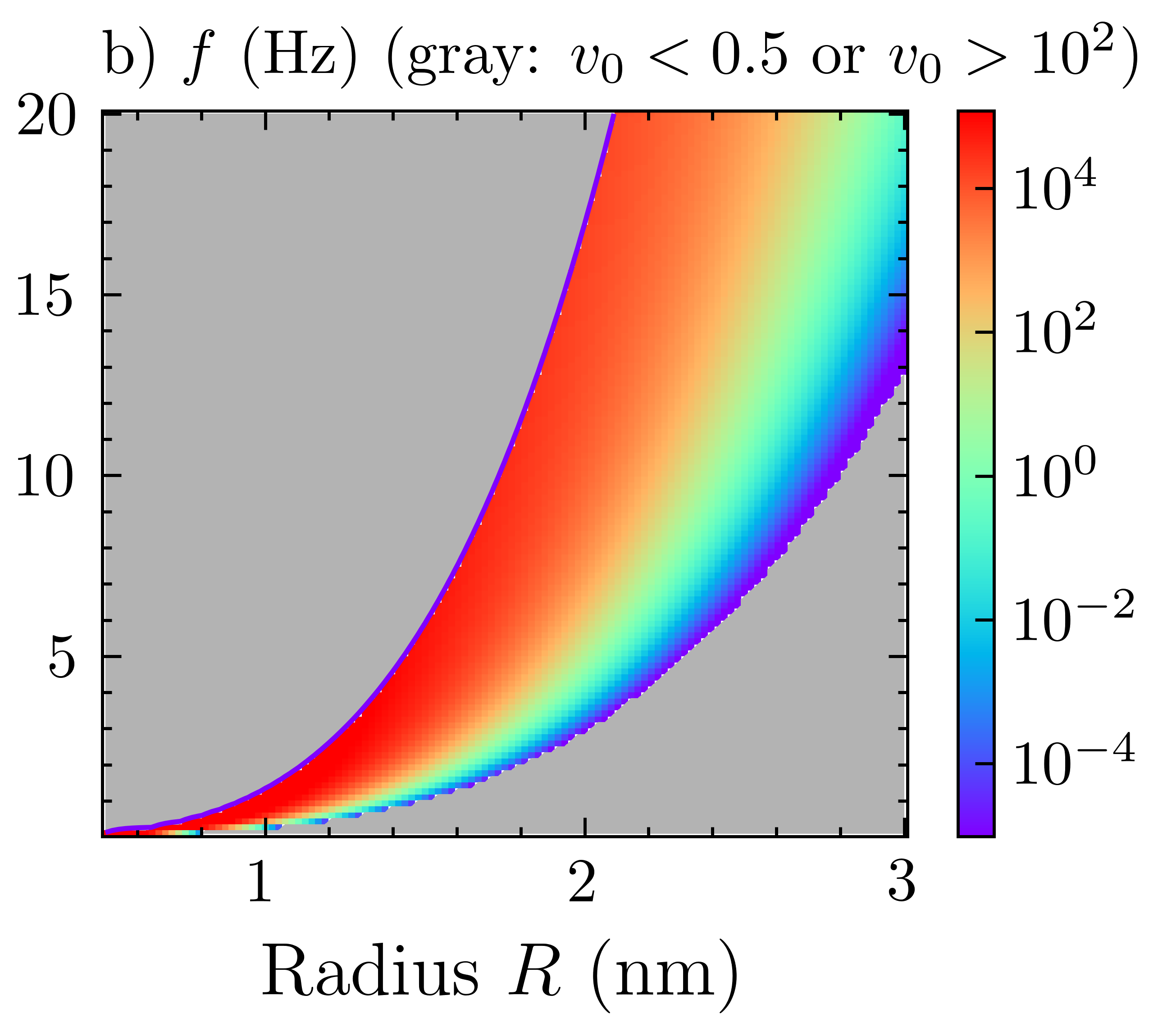}
  \includegraphics[width=0.521\linewidth]{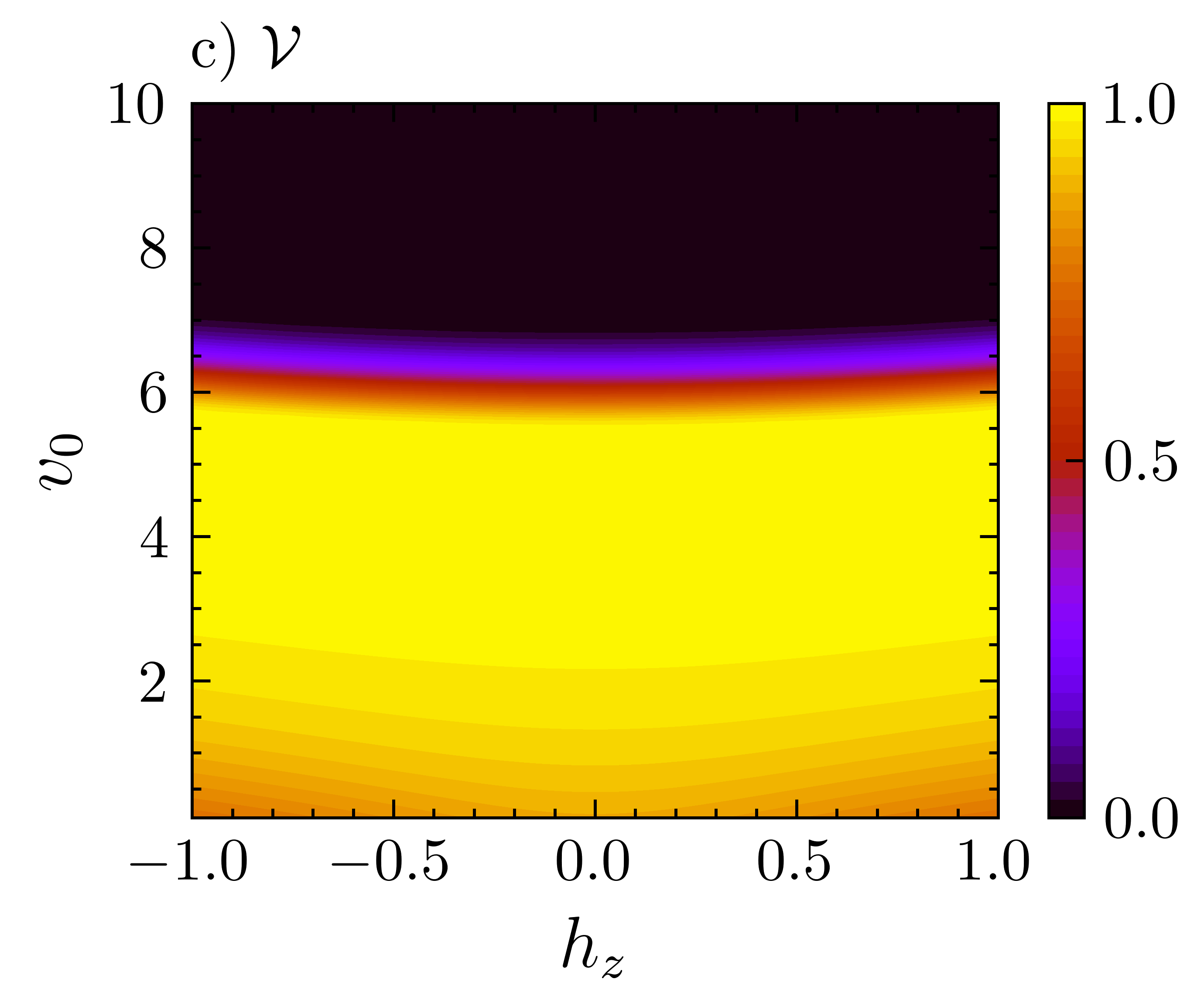}
  \includegraphics[width=0.479\linewidth]{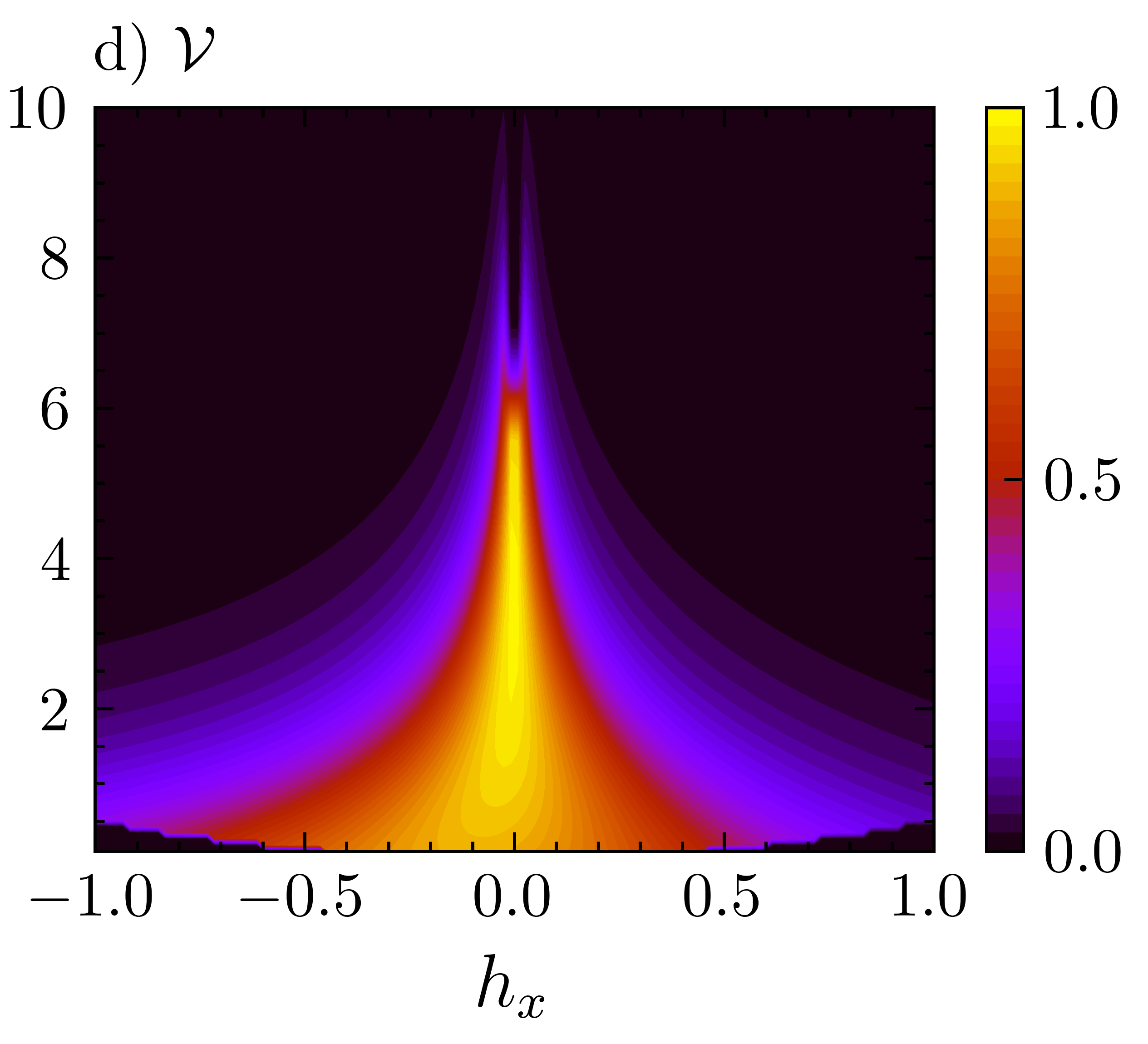}
  \includegraphics[width=0.52\linewidth]{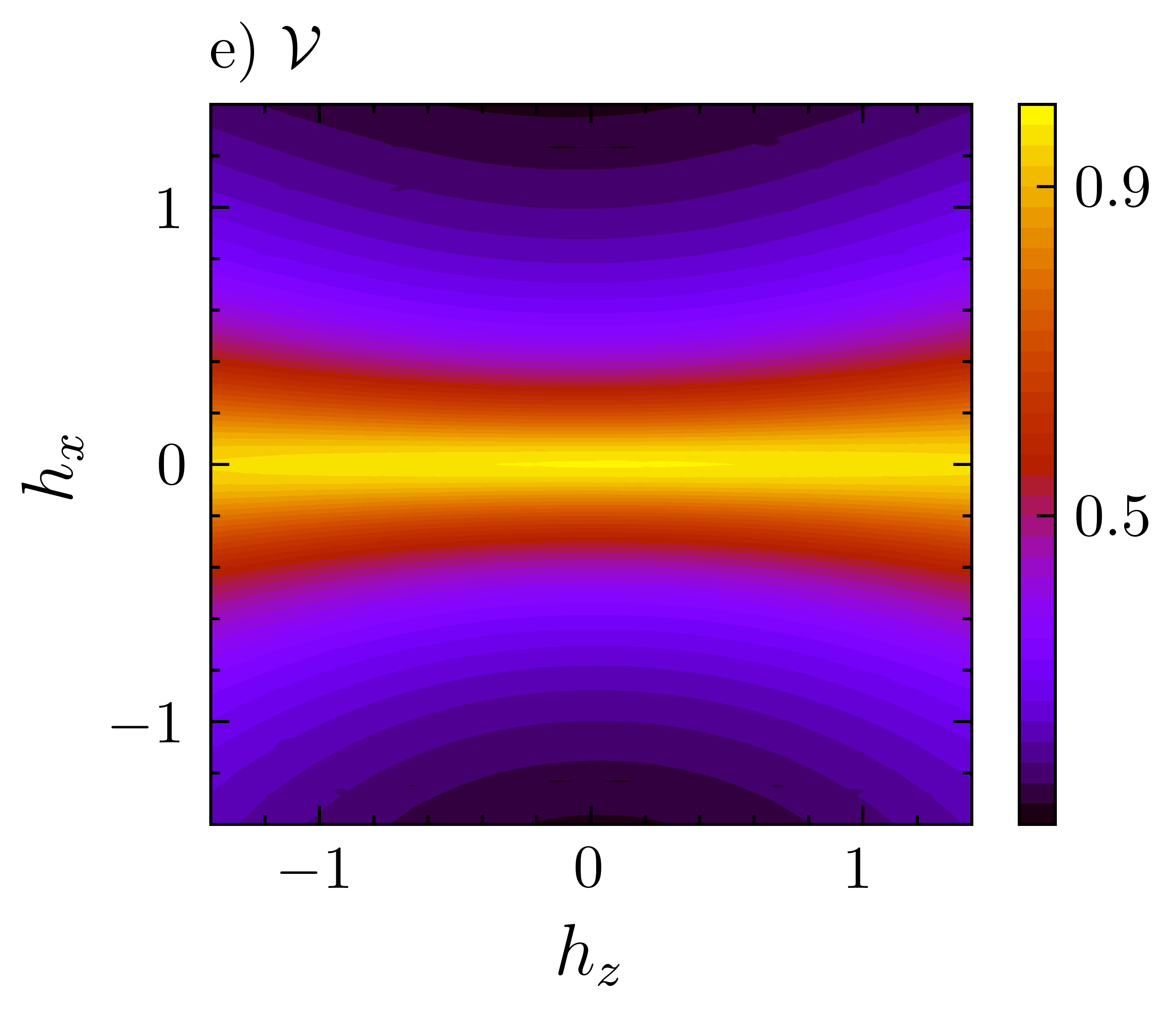}
  \includegraphics[width=0.465\linewidth]{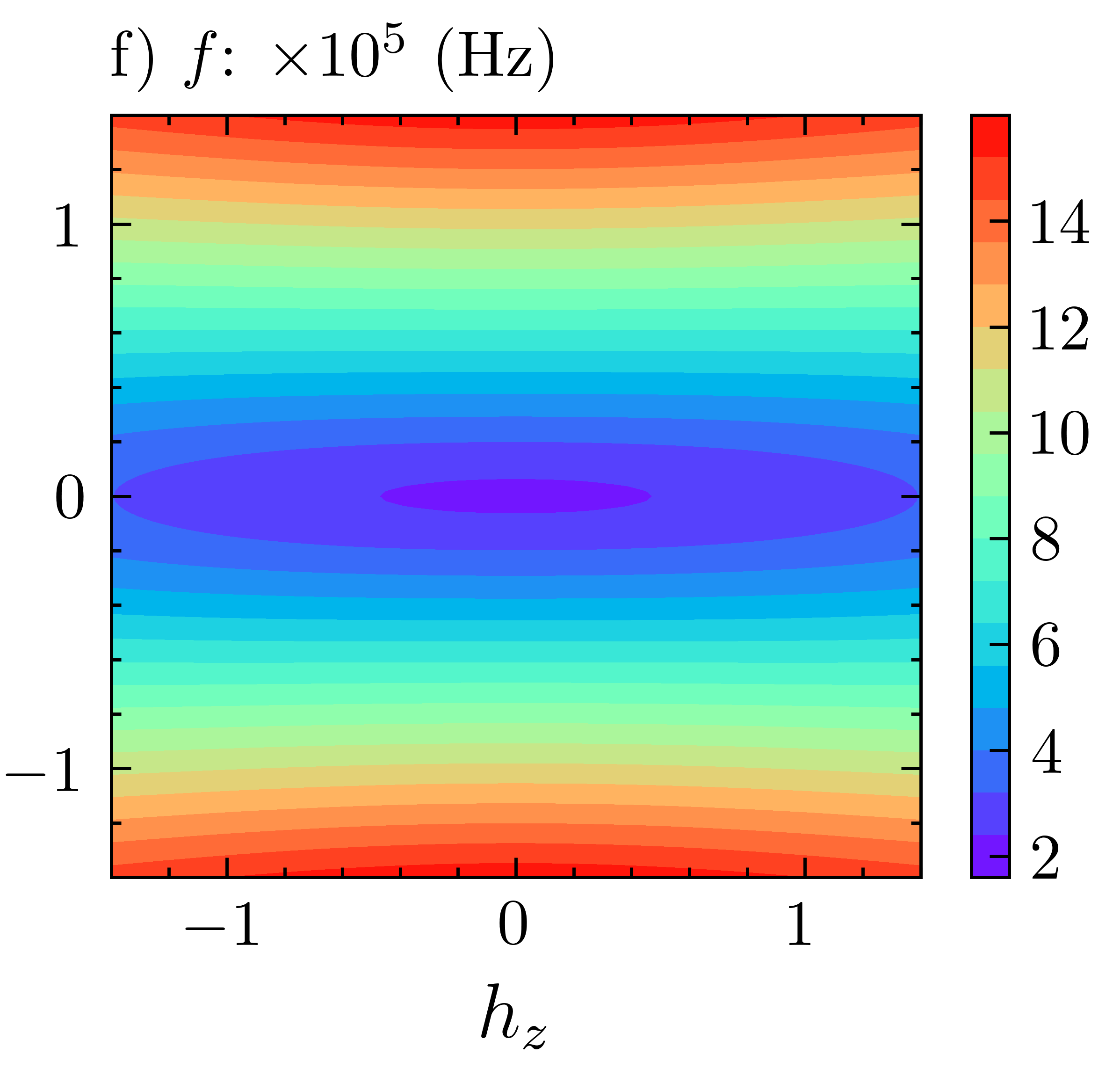}
  \caption{{Visibility and tunneling rate for the rotational double-well system across the relevant parameter space. Figs. (a) and~(b) show, respectively, the tunneling visibility and tunneling frequency as functions of particle radius $R$ and plate separation $L$ in the external field-free case $h_x=h_z=0$. Figs~(c)--(e) display visibility density plots in the planes $(v_0,h_z)$, $(v_0,h_x)$, and $(h_z,h_x)$, and ~(f) shows the corresponding tunneling frequency in the $(h_z,h_x)$ plane. These plots summarize the experimentally viable regime for rotational quantum tunneling, highlighting the trade-off between strong localization in the two wells, finite tunnel splitting, and high oscillation visibility. Gray regions denote parameter ranges outside the tunneling regime. The rest of the parameters are given in  Table \ref{tab:physical_correspondence}.}
} 
  \label{fig:vis1}
\end{figure}

In Fig.~\ref{fig:vis1} we provide parameter regimes in which rotational quantum tunneling should be observable in the superconducting trap. In the free field  case ($h_x = h_z = 0$),  Figs 5(a) and 5(b) show how the tunneling visibility and tunneling frequency depend on the particle radius $R$ and the plate separation $L$. The observable tunneling is confined to an intermediate window rather than occurring for all geometries. The bright yellow region in Fig.~\ref{fig:vis1}(a) indicates that high-amplitude oscillations are obtained only when the trap is neither too weak nor too strongly confining. In other words, the geometry must be chosen such that the double-well potential is strong enough to localize the rotor into two preferred orientations, but not so strong that the barrier suppresses tunneling almost completely. The gray regions therefore mark parameter ranges that are excluded from the intended tunneling regime in the current scenario, while the dark region on the right indicates parameter combinations for which the visibility of the oscillation between right and left well becomes very small. Fig.~\ref{fig:vis1}(b) complements this by showing the tunneling frequency on the same $(R,L)$ plane. The fastest oscillations occur toward the lower-left 
part of the allowed window, while the tunneling frequency decreases as one moves toward larger radius of the magnet and smaller plate separations. This behavior is physically reasonable, as once the particle becomes larger, its moment of inertia increases, and once the trap becomes smaller, the coherent transition between wells becomes slower. For the remainder of this work we choose the parameters in Table \ref{tab:physical_correspondence}.
Figs.~\ref{fig:vis1}(c) and \ref{fig:vis1}(d), shows how the visibility changes when the dimensionless barrier height parameter $v_0$ is combined with either $h_z$ or $h_x$ respectively. These plots are especially useful because they separate the effects of the two field directions. In Fig.~\ref{fig:vis1}(c), the visibility depends mainly on $v_0$, while the dependence on $h_z$ is comparatively weak and approximately symmetric. This reflects the fact that the $h_z$ term distorts the barrier shape but, at $h_x=0$, does not 
favor one well over the other. By contrast, Fig.~\ref{fig:vis1}(d) shows a much stronger sensitivity to $h_x$. Since the $h_x$ term breaks the left-right symmetry of the double well by making one orientation energetically preferred, increasing $|h_x|$ reduces the visibility of coherent oscillation. This contrast becomes even clearer in Figs. \ref{fig:vis1}(e) and \ref{fig:vis1}(f), where both field components are varied simultaneously. Fig.~\ref{fig:vis1}(e) shows a bright horizontal line centered around $h_x\approx 0$, extending over a broad interval of $h_z$. The interpretation is that the tunneling remains robust over a comparatively wide range of perpendicular field strengths, provided the symmetry-breaking component $h_x$ stays small. In other words, the system tolerates changes in $h_z$ much better than comparable changes in $h_x$. Fig.~\ref{fig:vis1}(f) then shows the corresponding tunneling frequency in the $(h_z,h_x)$ plane. The frequency is lowest near the unbiased central region and increases mainly with increasing $|h_x|$, whereas the dependence on $h_z$ is small. Taken together, Figs.~\ref{fig:vis1}(e) and \ref{fig:vis1}(f) highlight an important experimental trade-off. Parameters that increase the oscillation frequency do not automatically improve the visibility of tunneling, because the same symmetry-breaking field that speeds up the dynamics can simultaneously reduce the tunneling visibility.

\subsection{Symmetries and Selection Rules}\label{sec.sym}
In the external field free case $h_x=h_z=0$, $H_{\rm sys}$ has two symmetries: A $C_2$ symmetry, meaning invariance of $H_{\rm sys}$ under the translation $\mathcal{T}:\theta\mapsto \theta+\pi$ of the rotation angle, and a reflection symmetry $\mathcal{R}$, i.e.~$H_{\rm sys}$ commutes with the
reflection operator $\mathcal{R}:\theta\mapsto -\theta$.  
The energy eigenstates can therefore be chosen real and with definite parity
$t_m=\pm 1$ under translation and $r_m=\pm 1$ under reflection,
\begin{eqnarray}
  \mathcal{T}\ket{\psi_m} &=& t_m\ket{\psi_m}\\
  \mathcal{R}\ket{\psi_m} &=& r_m\ket{\psi_m}\,.
\end{eqnarray}
Possible perturbations $U$ of the Hamiltonian that we consider her, be
it because of $h_x$ or $h_z$ different from zero, or due to a coupling
$S^{(\alpha)}$ to the environment, are from the set $\{\cos\theta,\sin\theta,\cos 2\theta, \sin 2\theta, L_\theta \}$. They are all even or odd under the two symmetries, $\mathcal{T} U \mathcal{T}^\dagger=t_U U$, and $\mathcal{R} U \mathcal{R}^\dagger=r_U U$ with $t_U,r_U\in\{\pm\}$.  The specific values of $t_U$ and $r_U$ are summarized in Table \ref{tab:turu}.
\begin{table}
  \centering
  \begin{tabular}{c|c|c|c|c|c|}
    & $\cos\theta$     & $\sin\theta$     & $\cos 2\theta$     & $\sin
                                                                 2\theta$
    & $L_\theta$ \\
    \hline
    $\mathcal{T}$: $t_U$&   - & - & + & + & +\\
    \hline
    $\mathcal{R}$: $r_U$&   + & - & + & - & -
  \end{tabular}
  \caption{Symmetries under $\mathcal{T}$, $\mathcal{R}$ of different
    potentials $U$. These can arise due to  additional magnetic fields
    or as system coupling agents in the interaction Hamiltonian.  }
  \label{tab:turu}
\end{table}  
This implies for matrix elements of $U$ that
\begin{equation}
  \label{eq:psiUpsi}
  \bra{\psi_m}U\ket{\psi_n}=t_Ut_mt_n \bra{\psi_m}U\ket{\psi_n}=r_Ur_mr_n \bra{\psi_m}U\ket{\psi_n}\,.
\end{equation}
If $t_Ut_mt_n=-1$ or $r_Ur_mr_n=-1$, we get immediately
$\bra{\psi_m}U\ket{\psi_n}=0$, whereas no conclusion can be drawn from
$t_Ut_nt_n=1$ or $r_Ur_nr_n=1$. If $m=n$, hence $r_m=r_n$ and
$t_m=t_n$, we can conclude that diagonal matrix elements
$\bra{\psi_m}U\ket{\psi_m}$ vanish if $r_U=-1$ or $t_U=-1$, which is
the case for all $U$ in Table \ref{tab:turu} with the exception of
$U=\cos 2 \theta$. On the other hand, if $r_m=-r_n$ or  $t_m=-t_n$, we
have that $\bra{\psi_m}U\ket{\psi_n}=0$ if $r_U=+1$ or $t_U=+1$. From
Table \ref{tab:turu} we see that this is the case for all $U$ there
with the exception of $U=\sin\theta$.  In particular, not only
transitions in the tunneling split-ground states (i.e.~between states
$\psi_0$ and $\psi_1$) due to $C_2$-symmetry preserving $\cos 2\theta$
couplings to the environment are forbidden in first order perturbation
theory, but also transitions by
couplings to $\cos \theta$, based on the reflection symmetry.

If $h_z\ne 0$, the hindering potential aquires an additional $h_z
\cos\theta$ term, and the $C_2$ symmetry is broken.  However, the
reflection symmetry remains, and the conclusions about vanishing matrix
elements can be based on $\mathcal{R}$ alone.  We see that also in
this situation transitions between the tunneling split-ground states
due to both $\cos 2\theta$ couplings and $\cos \theta$ couplings are
forbidden in first order perturbation theory.

\begin{figure*}[t!]
  \centering
  \includegraphics[width=0.99\linewidth]{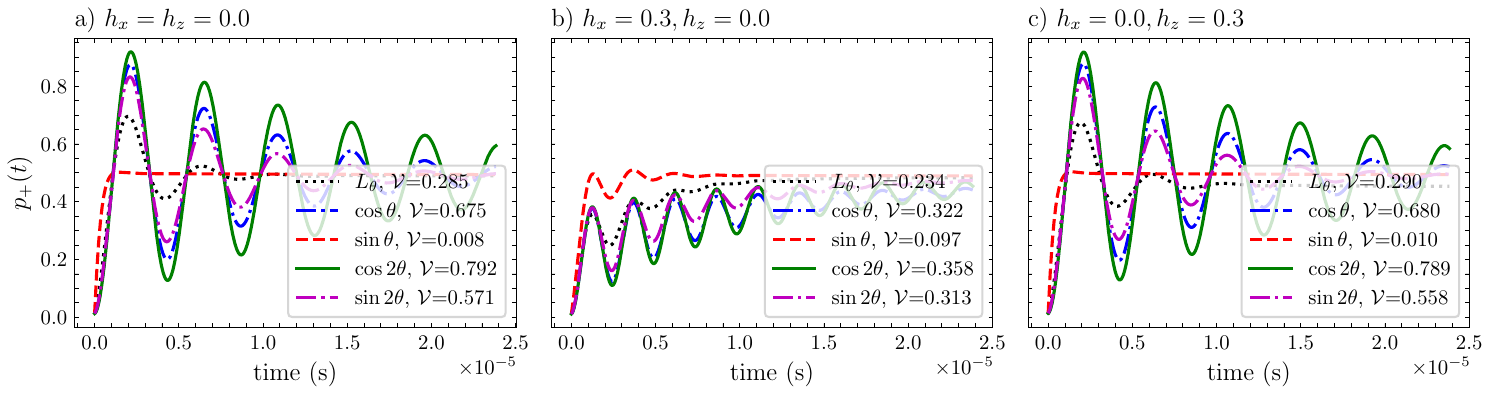}
  \\
    \includegraphics[width=1.0\linewidth]{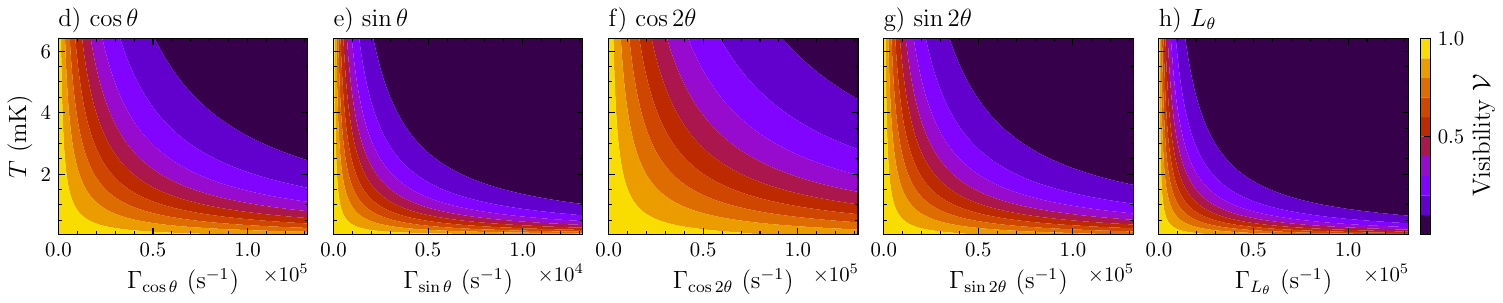}
  \caption{Decoherence of rotational tunneling for different system-environment coupling operators:  $\cos\theta$, $\sin\theta$, $\cos 2\theta$, $\sin 2\theta$, and $L_\theta$. Figs~(a) to (c) show the dissipative time evolution of the right-well  probability $p_+(t)$ for the coupling operators, for three different field configurations: (a) $h_x=h_z=0$, (b) $h_x=0.3 \rightarrow (3.169\times 10^{-8} \mathrm{T}),\ h_z=0$, and (c) $h_x=0,\ h_z=0.3 \rightarrow (3.169\times 10^{-8} \mathrm{T})$. The visibility values $\mathcal{V}$ calculated from the oscillations are indicated in the legends and temperature for all environments are chosen as $T = 3.2$\,mK and decay rates are $\Gamma_{\alpha} = 1.3 \times 10^4\mathrm{s}^{-1}$. 
  Figs~(d) to (h) show the corresponding tunneling visibility as a function of the effective decoherence rate $\Gamma_\alpha$ and temperature $T$ for the five coupling operators $\cos\theta$, $\sin\theta$, $\cos 2\theta$, $\sin 2\theta$, and $L_\theta$, respectively. The figure demonstrates that the loss of tunneling coherence depends strongly on the symmetry of the environmental coupling. Symmetry preserving channels, especially $\cos 2\theta$, retain high visibility over a much broader parameter range, whereas symmetry breaking channels like $\sin\theta$ suppress the oscillations more rapidly. The rest of the parameters are fixed and given in the Table \ref{tab:physical_correspondence}.
  }
  \label{fig:decoh}
\end{figure*}

\subsection{Decoherence effects for different couplings to the environment}
{In Fig.~\ref{fig:decoh} we show  how different environmental coupling channels suppress the coherent rotational tunneling of the trapped magnetic dipole. The initial states for these figures considered are a superposition of the lowest two levels to localize the particle in the left well. The upper row, Fig.~\ref{fig:decoh}~(a) to (c), shows the time evolution of the right-well probability $p_+(t)$ for five system environment coupling operators, $L_\theta$, $\cos\theta$, $\sin\theta$, $\cos 2\theta$, and $\sin 2\theta$, under three different additional field configurations. In all cases the oscillations decay toward a stationary value close to $p_+=1/2$, which is the expected signature of decoherence. The important point, however, is that the damping is strongly channel-dependent: the $\cos 2\theta$ coupling consistently preserves the largest visibility, while $\sin\theta$ and $L_\theta$ produce much stronger decoherence. This trend is most transparent in Fig.~\ref{fig:decoh}(a), corresponding to the fully symmetric case $h_x=h_z=0$. There, the visibility shows a clear hierarchy: $\cos 2\theta$ gives the largest visibility, followed by $\cos\theta$ and $\sin2\theta$, whereas $\sin\theta$ and $L_\theta$ are much more destructive. This agrees with the symmetry analysis. In the external field-free double well ($h_z=h_x=0$), the Hamiltonian has both $C_2$ symmetry and reflection symmetry, so matrix elements between the tunneling-split ground states vanish for symmetry-preserving channels such as $\cos 2\theta$, and are also suppressed for $\cos\theta$ by reflection symmetry. As a result, these couplings cannot distinguish well the two components of the tunneling doublet,
and the coherent oscillations survive longer. By contrast, the $\sin\theta$ and $L_\theta$ channels couple opposite-parity states. Fig.~\ref{fig:decoh}(c), where $h_z\neq 0$ but $h_x=0$, shows that this protection remains largely intact when a perpendicular field is present. Although the $h_z\cos\theta$ term breaks the $C_2$ symmetry, the reflection symmetry is still preserved, which is enough to maintain the suppression of direct transitions between the two lowest tunneling states for both $\cos\theta$ and $\cos 2\theta$. This is why the ordering of the visibility values in Fig.~\ref{fig:decoh}(c) is very similar to Fig.~\ref{fig:decoh}(a). The $\cos 2\theta$ and $\cos\theta$ couplings remain those with the weakest damping of the tunneling motion, and the symmetry-breaking channels continue to damp the oscillations more strongly. Physically, the figure shows that a  magnetic field perpendicular to the superconducting plates deforms the double well without fully destroying the symmetry protection of the tunneling subspace. The behavior changes more noticeably in Fig.~\ref{fig:decoh}(b), where $h_x\neq 0$ and $h_z=0$. The $h_x\sin\theta$ term makes the double well asymmetric and breaks the reflection symmetry that protected  
the tunneling 
in the previous two cases. Consequently, the separation between the different curves becomes less pronounced and the advantage of the symmetry-preserving channels is reduced. And, $\cos 2\theta$ still yields the largest visibility among the five 
couplings to the environment, indicating that it 
still preserves best the coherent tunneling. Fig.~\ref{fig:decoh} therefore illustrates an important qualitative point. A perpendicular magnetic field $h_z$ can still leave substantial symmetry protection, while a parallel magnetic field $h_x$ spoils more directly  the symmetry structure that 
protects coherent tunneling.}

Figs.~\ref{fig:decoh}(d)--(h) summarize the same phenomena in a more global way by plotting the tunneling visibility $\mathcal{V}$ as a function of the decoherence rates $\Gamma_\alpha$ and the bath temperature $T$ for each coupling operator separately. In every figure the visibility is highest in the lower-left corner, where both $\Gamma_\alpha$ and $T$ are small, and decreases monotonically as either parameter is increased. This confirms the expected competition between coherent tunneling and environment-induced relaxation or dephasing: stronger coupling to the bath or larger thermal occupation both enhance the 
decoherence. At the same time, the shapes of the colored regions reveal that the channels are not equally damaging. The $\cos 2\theta$ shows a  high-visibility region over the largest range of parameters, while $\cos\theta$ is the next most robust. The $\sin\theta$ coupling is clearly the most  
detrimental one.  The horizontal axis with the decay rate $\Gamma_{\sin\theta}$ is plotted on a smaller scale than for the other couplings, indicating that substantial visibility loss already occurs for significantly smaller decoherence rates. The $\sin 2\theta$ and $L_\theta$ channels lie in between, but are still less favorable than $\cos 2\theta$.

{Environmental couplings that respect the symmetry structure of the hindering potential have a much weaker effect on the tunneling doublet, because they cannot efficiently distinguish the two localized orientations that form the coherent superposition. In contrast, symmetry-breaking couplings act as which-well probes and therefore destroy the oscillations much more rapidly. Fig.~\ref{fig:decoh} thus gives direct numerical support to the idea of a symmetry-protected or decoherence-free tunneling subspace for the rotational motion of the levitated magnetic dipole.}

\subsection{Physical Decoherence Mechanisms} \label{sec:dec}

We derive the following physical rates for different environment effect on the rotational tunneling.

\paragraph{Gas scattering (rotational localization).}

Residual gas molecules scatter off the nanoparticle and carry away
information about its angular orientation.  For a particle that is not
perfectly rotationally symmetric, this imprints a which-orientation
record in the environment and induces angular decoherence.  The
microscopic scattering model of Carlesso \textit{et
  al.}~\cite{carlesso_perturbative_2021}, adapted to a free planar rotor
with surface roughness (Appendix~\ref{app:scattering}), yields a
Lindblad master equation of the \emph{pure-localization}
type: in the
angle representation, the off-diagonal elements of the density matrix
decay as
\begin{equation}
  \dot\rho(\theta,\theta')\big|_{\rm gas}
  = -\Lambda_{\rm R}\,
    \sin^2\!\!\left(\frac{\theta-\theta'}{2}\right)\rho(\theta,\theta'),
  \label{eq:loc_kernel_main}
\end{equation}
where $\Lambda_{\rm R}$ (units~s$^{-1}$) is the localization rate,
determined by the gas density, temperature, and particle roughness
(Appendix~\ref{app:scattering}).  
The localization kernel
$\sin^2\!\!\left((\theta-\theta')/2\right)$ is diagonal in the
orientation basis: it suppresses coherences between distinct orientations. 
It corresponds to the heavy-particle limit ($m_{\rm
  gas}/M\to 0$) of the quantum linear Boltzmann equation for
rotational degrees of
freedom~\cite{stickler_spatio-orientational_2016,papendell_quantum_2017},
in which the particles collisions lead to small angular-momentum kicks of the rotor and hence a slow diffusion of its angular momentum.
In this so-called monitoring approach, the orientation of the rotor is only imprinted in the environment in the collisions, and hence leads to decoherence. In this approach the orientation of the particle only  enters parametrically into the scattering amplitudes. In the case of a planar rotor, this leads to a Lindblad master equation with angular momentum ladder operators as Lindblad operators \cite{stickler_rotational_2018}.  
For rotors trapped additionally in a hindering potential one expects the orientation of the rotor to be even less modified by a single impact of a particle, and hence the master equation derived in \cite{stickler_spatio-orientational_2016} still to be valid.

The localization kernel~\eqref{eq:loc_kernel_main} can be recast
as a pair of Lindblad channels.  Using the identity
$\e^{i(\theta-\theta')}+\e^{-i(\theta-\theta')}-2
=-4\sin^2[(\theta-\theta')/2]$, one verifies that
\begin{equation}
  \dot\rho\big|_{\rm gas}
  = \frac{\Lambda_{\rm R}}{4}
    \Bigl(\mathcal{D}[\e^{i\theta}]
    +\mathcal{D}[\e^{-i\theta}]\Bigr)\rho,
  \label{eq:gas_Lindblad}
\end{equation}
where $\mathcal{D}[L]\rho = L\rho L^\dagger - \tfrac{1}{2}\{L^\dagger
L,\rho\}$. Here, $e^{\pm i\theta}$ denotes the unitary angular-momentum ladder operator, $e^{i\theta}\ket{m} = \ket{m+1}$, which acts as a shift operator on the angular-momentum eigenstates (see also Appendix~A). The two jump operators
\begin{equation}
  L_{\pm} = \sqrt{\tfrac{\Lambda_{\rm R}}{4}}\;\e^{\pm i\theta}
  \label{eq:gas_jumps}
\end{equation}
shift the angular-momentum quantum number by $\pm 1$ and are the
rotational analogues of the position-localization operators in
translational collisional decoherence.

Writing $e^{i\theta}=\sum_{m,n}\langle\psi_m|e^{i\theta}|\psi_n\rangle\,|\psi_m\rangle\langle\psi_n|$ and evaluating the dissipator $\mathcal{D}[e^{i\theta}]\rho$ directly (e.g. retaining the \emph{full} operator $e^{i\theta}$ as the jump operator without the rotating wave approximation (RWA) that normally leads to a decomposition into individual Bohr-frequency channels) yields a Lindblad equation that is exact within the pure-localization model and is already of Lindblad (GKLS) form,  as required by the Lindblad theorem for Markovian, trace-preserving and positivity preserving generators of a semi-group. The RWA, which eliminates cross terms between jump operators of different Bohr frequencies and is essential in the standard Born--Markov approximation for a bosonic bath, is not needed here because the localization kernel is derived independently from the scattering theory and is \emph{ab initio} in Lindblad form. When $e^{i\theta}$ is expanded in the energy eigenbasis, the resulting master equation contains cross terms between pairs of energy eigenstates with different Bohr frequencies; these terms are retained because the dissipator is already in Lindblad form and no secular approximation is applied.

\begin{figure*}[t!]
  \centering
  \includegraphics[width=0.49\linewidth]{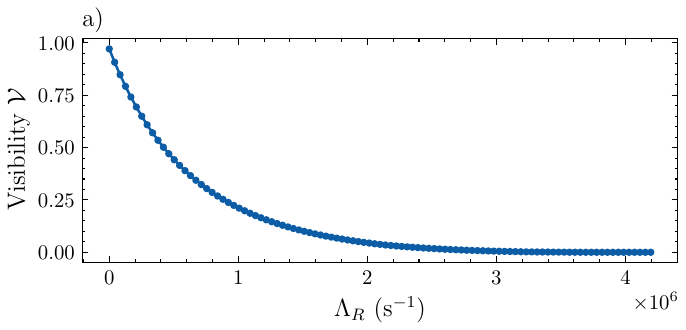}
    \includegraphics[width=0.49\linewidth]{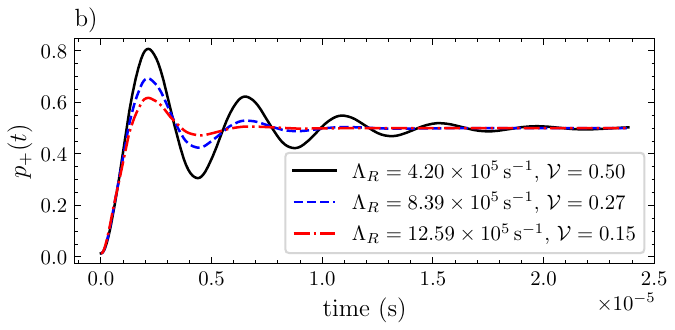}
\caption{Gas-scattering decoherence of rotational tunneling in the field-free case ($h_x = h_z = 0$), using the pure-localization dissipator with jump operators $L_{\pm}$. (a)~Tunneling visibility as a function of the localization rate $\Lambda_{\mathrm{R}}$. (b)~Time evolution of the right-well probability $p_+(t)$ for three values of $\Lambda_{\mathrm{R}}$, with corresponding visibilities indicated in the legend. The remaining parameters are given in Table~\ref{tab:physical_correspondence}.}
  \label{fig:decoheq21}
\end{figure*}

Two features of the pure-localization dissipator~\eqref{eq:gas_Lindblad} should be noted. First, because it originates from impulsive collisions rather than from coupling to a bath of harmonic oscillators, there is no underlying spectral density $J(\omega)$; the single parameter $\Lambda_{\rm R}$ encodes the full strength of the gas-scattering channel. Second, because the jump operators $e^{\pm i\theta}$ enter with equal rates, the dissipator is symmetric under $\theta\to-\theta$. In the angular-momentum basis, angle-localization acts as a diffusion process that spreads the angular-momentum distribution, driving populations toward the uniform distribution rather than the Gibbs state at the physical gas temperature $T$.

For the tunneling doublet this is inconsequential, since the splitting $\Delta$ is exponentially small and any realistic temperature satisfies $k_{\rm   B}T\gg\hbar\omega_{01}$, so that thermal equilibrium within the doublet is effectively uniform.  

For the parameters used here (helium background gas, $T=3.2\,\mathrm{mK}$, $n\sim 10^{11}\,\mathrm{m}^{-3}$, and the roughness model of Appendix~\ref{app:scattering}), we obtain
\begin{eqnarray}
    \Lambda_\mathrm{R} \sim 
    9.7\times10^{-6}\,\text{s}^{-1}.
\end{eqnarray}
giving an effective GKLS coupling strength 
 $\Gamma_{\rm gas}=\Lambda_{\rm R}=
 9.7\times10^{-6}\,\text{s}^{-1}$. 
Both the generic ($\cos\theta$) and the $C_2$-symmetric ($\cos 2\theta$) gas-scattering implementations are included in the numerical analysis.

In Fig.~\ref{fig:decoheq21} we show the effect of residual-gas  scattering on the tunneling coherence, using the  pure-localization dissipator of Eqs.~\eqref{eq:gas_Lindblad}--\eqref{eq:gas_jumps}. The visibility decreases  monotonically with the localization rate $\Lambda_{\mathrm{R}}$  and drops below $50\%$ when $\Lambda_{\mathrm{R}}$ becomes  comparable to the tunneling frequency $f_T \sim 10^5\,$Hz. 
The tunneling oscillations therefore survive for roughly $10^{11}$ cycles before  gas-scattering decoherence becomes appreciable, confirming that  this channel does not limit the observability of rotational  quantum tunneling under ultra-high-vacuum conditions.

In the limit of a perfectly spherically symmetric particle, the surface roughness vanishes ($\Delta r / r \to 0$) and  with it the localization rate $\Lambda_{\mathrm{R}} \to 0$,  since scattered gas molecules can no longer acquire  which-orientation information. The tunneling-split ground  states then span a decoherence-free subspace with respect  to the gas-scattering channel: the environment cannot  distinguish the two orientations and coherence is preserved indefinitely. For a particle with $C_2$-symmetric roughness,  this protection is partial, the dominant jump operators  $e^{\pm 2i\theta}$ cannot connect the tunneling doublet  due to opposite $C_2$ parity, and decoherence enters only  through higher-order angular harmonics of the surface profile.

\paragraph{Eddy currents} Eddy-currents are induced by the time-varying magnetic field of the rotating nano-particle both in the superconducting plates and in the nano-particle itself. The corresponding dissipation is modeled by the system coupling operator $ S^{(\rm eddy)}=\hat L_\theta , $  with the corresponding GKLS jump operators constructed in the energy eigenbasis as in Eq.~\eqref{eq:jumps_main_energybasis}. For the parameter range considered here, however, the microscopic estimates of Appendix~\ref{app:eddy} show that this channel 
is negligible and can be dropped. For dissipation in the superconducting plates gives the scale for the damping rate
\begin{equation}
\Gamma^{\mathrm{eddy,s}} \approx 1.07\times 10^{5}\,\frac{n_n}{n}\ \mathrm{s^{-1}}.
\label{eq:gamma_eddy_sc_main}
\end{equation}
where $n_n/n$ is the normal-fluid (quasiparticle) fraction in the two-fluid description. For Ta ($T_c\simeq 4\,{\rm K}$) and our operating temperature $T\simeq 10^{-3}\,{\rm K}$,  
the thermal quasiparticle fraction is exponentially small ($\propto e^{-\Delta/k_BT}$ with gap $\Delta(0)\simeq 1.76\,k_B T_c$), so Eq.~\eqref{eq:gamma_eddy_sc_main} is effectively zero on all timescales relevant to the tunneling dynamics. For eddy-current losses inside the particle, the estimate in Appendix~\ref{app:eddy} is even smaller,
\begin{equation}
\Gamma^{\mathrm{eddy},\,p}=
\frac{2\pi\mu_0^2\zeta(3)^2}{135}\,
\frac{M_s^2\sigma R^{11}}{IL^6}
\sim 10^{-11}\,\mathrm{s}^{-1},
\end{equation}
for the nominal nanometer-scale parameters used there. Both eddy-current contributions are therefore many orders of magnitude below the gas-scattering localization rates and the tunneling frequencies of interest, and we neglect this channel in the numerical simulations.

\paragraph{Readout back-action.}
 
A second source of decoherence, linked to the magnetic field, is the back-action from coupling the dipole to a readout device such as a SQUID.  A pickup coil with its normal parallel to the superconducting plates produces a $\sin\theta$ Lindblad operator; a perpendicular orientation gives a $\cos\theta$ operator, which benefits from reflection-symmetry protection of the tunneling doublet 
(cf.\ Table~\ref{tab:turu}). Adiabatic elimination of the SQUID degrees of freedom~\cite{Mueller2025} (Appendix~\ref{app:readout}) yields the decoherence rate
\begin{equation}
  \Gamma^{\mathrm{SQ}}_{\sin\theta}\sim 10^{-35}\,\mathrm{s}^{-1},
\end{equation}
strongly suppressed by the large frequency mismatch between the SQUID ($\sim\!10\,$GHz) and the rotor ($\sim\!10^5\,$Hz).  The SQUID readout therefore introduces no measurable back-action on the tunneling dynamics.

\paragraph{Phonons} 
Acoustic phonons arise from elastic vibrations of the superconducting trap, which modulate the plate separation and therefore the orientational trapping potential of the nanomagnet. This couples the rotor to the phonon bath through the operator $\cos 2\theta$, so that acoustic noise preserves the symmetry of the double-well potential. Consequently, in the symmetric trap the direct transition between the two lowest tunnelling states is forbidden to first order, and the dominant phonon-induced decoherence can occur only through weaker couplings to higher excited states. The corresponding microscopic acoustic spectral density is 
{superohmic}, $J_{\mathrm{ac}}(\omega)\propto \omega^3$, with a prefactor set by the trap geometry, the magnetic energy scale, and the elastic properties of the superconducting plates (see Appendix \ref{app:phonon}). For the parameters considered here, this leads to an extremely small  
decay rate, 
on the order of 
\begin{equation}
    \Gamma_{\cos{2\theta}}^{\mathrm{ac}} = 7.714\times10^{-24} \,\mathrm{s}^{-1}
\end{equation}
at mK temperatures. This shows that coupling to acoustic phonons is negligible on the timescales relevant for tunneling dynamics.

\paragraph{Seismic noise} 

Classical vibrations of the trap center at frequencies $1-10^3\,\mathrm{Hz}$ couple quadratically to the rotor through $\cos^2\!\theta\,\delta z^2$, with decay rates that scale as
\begin{eqnarray}
     \Gamma^{\mathrm{vib}}_{\cos 2\theta} \propto (\sqrt{S_z})^4,
\end{eqnarray}
 where $\sqrt{S_z}$ is the displacement amplitude spectral density of the platform. This quartic scaling makes vibration isolation the key experimental control knob: For a standard cryostat ($\sqrt{S_z} \sim 10^{-11}\,\mathrm{m}/\!\sqrt{\mathrm{Hz}}$, ~\cite{NUCLEUS2025}) one finds 
 decoherence rates $\gamma_\uparrow(\omega_T),\gamma_\downarrow(\omega_\text{T})\sim 
10^4$\,s$^{-1}$ 
whereas state-of-the-art passive isolation ($\sqrt{S_z} \sim 10^{-13}\,\mathrm{m}/\!\sqrt{\mathrm{Hz}}$, demonstrated by de~Wit \textit{et~al.}~\cite{deWit2019}) pushes 
them to $\sim 10^{-4}$\,s$^{-1}$. 

\paragraph{Comparison}\label{sec.comp_dec}
Among the mechanisms considered, vibration-induced decoherence from seismic noise is the most severe if left unmitigated. After seismic noise, the second most dominant channel is gas scattering. Gas particles scatter off of the nanoparticle's surface and (should the particle not be perfectly spherical) obtain information about the particle orientation. For a particle roughness of approximately $\sim5\%$, the decoherence rate is still ten orders of magnitude below our tunneling rate.

\section{Conclusion}

We have investigated the rotational quantum tunneling dynamics of a magnetic dipole levitated between two parallel superconducting plates, including a comprehensive analysis of the principal decoherence mechanisms. The trapping potential, derived from the method of image dipoles, yields a periodic double-well structure with $C_2$ symmetry, whose barrier height can be further controlled by applied magnetic fields parallel or perpendicular to the plates.

By systematically estimating and comparing the decoherence rates from acoustic phonons, seismic vibrations, eddy currents, and residual gas scattering, we find that seismic vibrations are the dominant decoherence mechanism if unmitigated (decoherence rates
$\gamma_\uparrow(\omega_T),\gamma_\downarrow(\omega_\text{T})\sim 
10^4$\,s$^{-1}$ for a standard cryostat). However, they can be brought under control through state-of-the-art passive isolation, exploiting the favorable quartic scaling $
\gamma_\uparrow(\omega_\text{T}),\gamma_\downarrow(\omega_T)\propto(\sqrt{S_z})^4$. Gas scattering becomes the dominant decoherence mechanism at millikelvin temperatures once vibration isolation of the experimental platform 
reduces the seismic-noise decoherence rate below $10^{-5}\,\text{s}^{-1}$. 
Acoustic phonon decoherence produces rates of order $10^{-24}\,\text{s}^{-1}$, which is entirely negligible. Eddy-current damping, both in the particle and in the superconducting plates, is exponentially suppressed at millikelvin temperatures and can be safely ignored. 

A central result of this work is the identification of a decoherence-free subspace protecting the rotational tunneling coherence, analogous to the symmetry protection observed in methyl-group tunneling in molecular solids. For a nanoparticle with a $C_2$-symmetric roughness profile, the gas-scattering jump operators ($e^{\pm 2i\theta}$) cannot connect the tunneling-split ground states, which carry opposite $C_2$ parity. Furthermore, the reflection symmetry $\mathcal{R}:\theta\to-\theta$ of the trapping potential forbids transitions between the ground-state doublet for all coupling operators {studied} except $\sin\theta$. This symmetry protection pushes the onset of decoherence to temperatures where thermally activated transitions to higher librational levels become significant. 
For a nanoparticle of radius $R=1\,$nm levitated in a trap of width $L = 84\,$nm, we predict a tunneling frequency of approximately $2.29\times10^5\,$Hz, and seismic noise decoherence rates on the order of $10^{-4}/\,$s for excellent vibration isolation.
The tunneling oscillations should therefore persist for 
$\sim 10^9 $ cycles before decoherence becomes appreciable, 
with visibility $\mathcal{V}$ well above the 10\% threshold. Even with good passive vibration isolation, on the order of $ 10^4 $ tunneling cycles should be observable

These findings outline a concrete experimental path toward observing rotational quantum tunneling of a mesoscopic object. The principal requirements are: (i) fabrication of single-domain magnetic nanoparticles with radii in the 1--3\,nm range and controlled surface quality; (ii) a superconducting trap with plate separation $\sim 50$--$200\,$nm, cooled to millikelvin temperatures in a dilution refrigerator; (iii) vibration isolation achieving $\sqrt{S_z}\lesssim 10^{-12}\,$m$/\sqrt{\mathrm{Hz}}$; and (iv) ultra-high vacuum within the trapping region. 

\textit{Acknowledgements --} 
FJH and DB acknowledge the EU EIC Pathfinder project QuCoM (101046973). FM further acknowledges the support from the Czech Science Foundation (Junior Star 25-17250M). TF and HU acknowledge funding from the EU Horizon Europe EIC Pathfinder project QuCoM (10032223), from the UK funding agency EPSRC (grants  EP/V035975/1, EP/V000624/1, EP/W007444/1, EP/X009491/1), and from the Leverhulme Trust (RPG-2022-57), as well as support from the QuantERA II Programme (project LEMAQUME) that has received funding from the European Union’s Horizon 2020 research and innovation programme under Grant Agreement No 101017733. EK acknowledges the project PID2023-152724NA-I00, with funding from MCIU/AEI / 10.13039/501100011033 and FSE+, by the project CNS2024-154818 with funding from MICIU/AEI /10.13039/501100011033.

\bibliography{refs.bib}
\onecolumngrid

\newpage
\appendix

\begin{table}[htbp]
\centering
\caption{Physical correspondence for the parameters considering $\mathrm{Nd}_{2}\mathrm{Fe}_{14}\mathrm{B}$  
}
\label{tab:physical_correspondence}
\begin{tabular}{lll}
\hline
\textbf{Quantity} & \textbf{Physical Value} & \textbf{Dimensionless} \\
\hline
Radius, $R$ & $1.0\times10^{-9}$ m&  \\
Length, $L$ & $8.4\times10^{-8}$ m&  \\
Density, $\rho$ & $7.5\times10^{3}$ kg\,m$^{-3}$&  \\
Saturation magnetization, $M_s$ & $1.0\times10^{6}$A\,m$^{-1}$  & \\
Moment of inertia, $I$ & $1.257\times10^{-41}$ kg\,m$^{2}$  & \\
Kinetic energy, $E_{\mathrm{k 
}}$ & $4.425\times10^{-28}$  J & \\
Corresponding rate, $E_\mathrm{k}/\hbar$ & $4.196\times10^{6}\ \mathrm{s}^{-1}$ & $1.0$ \\
Barrier height, $V_0$ & $1.779\times10^{-27}$  J & $v_0$ = 4.02089\\
Lowest doublet splitting, $\Delta E$ & $1.518\times10^{-28}$  J  & \\
Tunneling frequency, $f_T$ & $2.290\times10^{5}$  Hz & \\
\hline
\end{tabular}
\end{table}

\section{System Hamiltonian in $|n\rangle$ basis}
We want to derive the Lindblad master equation for the quantum rotor
in the energy eigenebasis $\ket{\psi_n}$. For this, we
  express the system coupling agents in the 
  angular momentum eigen basis $\{ |m\rangle \}$, where $m \in
  \mathbb{Z}$ labels eigenstates of angular momentum operator
  $L_\theta$, $L_\theta\ket{m}=m\ket{m}$.  We derive the matrix form of the rotor Hamiltonian in angular momentum eigen basis, starting from the continuous operator in $\theta$-space. We work with the angular coordinate $\theta\in[-\pi,\pi)$ and define angular momentum eigen states as
\begin{equation}
| n \rangle = \int_{-\pi}^{\pi}d\theta |\theta\rangle \langle \theta | n \rangle, \quad \mathrm{with} \quad 
\langle \theta| n \rangle = \frac{1}{\sqrt{2\pi}} e^{i n \theta},
\qquad n\in\mathbb{Z},
\label{eq:A1}
\end{equation}
which diagonalizes the kinetic-energy operator,
\begin{equation}
-\frac{\hbar^2}{2I}\frac{\partial^2}{\partial\theta^2}|n\rangle
= \frac{\hbar^2 n^2}{2I}|n\rangle .
\label{eq:A2}
\end{equation}
The rotor Hamiltonian is given by 
\begin{equation}
\hat H_{\rm sys}
=
-\frac{\hbar^2}{2I}\frac{\partial^2}{\partial\theta^2}
+\frac{V_0}{2}\bigl(1+\cos 2\theta\bigr)
-\mu B_z \cos\theta
-\mu B_x \sin\theta ,
\label{eq:A3}
\end{equation}
where $V_0$ sets the double-well barrier height in the absence of magnetic fields and $B_z,B_x$ are external magnetic fields coupling to the magnetic dipole moment $\mu$.
In the symmetric case $B_x=B_z=0$, the potential has degenerate minima at $\theta=\pm\pi/2$. Using the identities
\begin{align}
\,\langle n | \cos(2\theta) | m \rangle &= \frac{1}{2} (\delta_{n,m+2} + \delta_{n,m-2})\\
\langle n|\cos\theta|m\rangle &= \tfrac12\bigl(\delta_{n,m+1}+\delta_{n,m-1}\bigr), \nonumber\\
\langle n|\sin\theta|m\rangle &= \tfrac{1}{2i}\bigl(\delta_{n,m+1}-\delta_{n,m-1}\bigr),
\label{eq:A4}
\end{align}
the Hamiltonian matrix elements $H_{nm}=\langle n|\hat H_{\rm sys}|m\rangle$ are
\begin{equation}
\begin{aligned}
H_{nm}
&=
\left(\frac{\hbar^2 n^2}{2I}+\frac{V_0}{2}\right)\delta_{nm}
+\frac{V_0}{4}\bigl(\delta_{n,m+2}+\delta_{n,m-2}\bigr) \\
&\quad
-\frac{\mu B_z}{2}\bigl(\delta_{n,m+1}+\delta_{n,m-1}\bigr)
-\frac{\mu B_x}{2i}\bigl(\delta_{n,m+1}-\delta_{n,m-1}\bigr).
\end{aligned}
\label{eq:A5}
\end{equation}
Equivalently, the Hamiltonian can be written in operator form as
\begin{equation}
\begin{aligned}
\hat H_{\rm sys}
&=
\sum_n\left(\frac{\hbar^2 n^2}{2I}+\frac{V_0}{2}\right)|n\rangle\langle n|
+\frac{V_0}{4}\sum_n\bigl(|n\rangle\langle n+2|+\mathrm{h.c.}\bigr) \\
&\quad
-\frac{\mu B_z}{2}\sum_n\bigl(|n\rangle\langle n+1|+\mathrm{h.c.}\bigr)
-\frac{\mu B_x}{2i}\sum_n\bigl(|n\rangle\langle n+1|-\mathrm{h.c.}\bigr).
\end{aligned}
\label{eq:A6}
\end{equation}

\begin{figure}[h]
  \centering
  \includegraphics[width=0.5\linewidth]{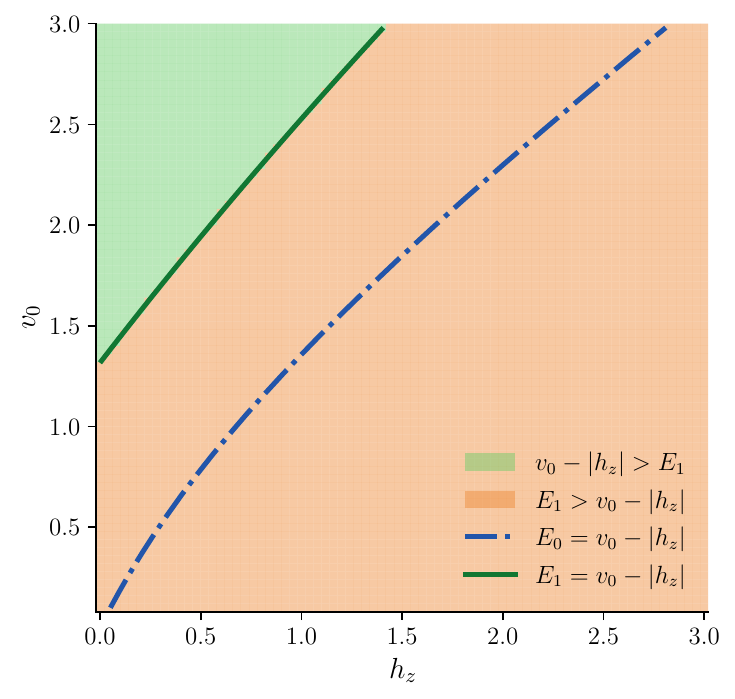}
  \caption{Boundary of the tunneling regime in the $(h_z, v_0)$ 
plane at $h_x = 0$. The solid green curve shows where the the effective barrier height $v_0-|h_z|$ is equal to the 
$E_1$ energy level, such that for the green shaded area, classical tunneling is not possible. In the green shaded region, both
$E_0$ and $E_1$ lie below the barrier and coherent tunneling between the two orientational wells is possible. Below the green curve in the orange shaded area, the first excited state can classically surmount the barrier and the system exits the tunneling regime.}

  \label{fig:crossing_analysis}
\end{figure}

\section{Microscopic master equation for the rotational tunneling}
\label{app:master}

For completeness, we sketch here the derivation of the Markovian master equation for the
rotational tunneling dynamics starting from a microscopic
system--bath model following \cite{Breuer06}. The derivation is carried out in full generality,
without restricting the interaction, and shows explicitly that
the Lindblad jump operators are defined in the energy eigenbasis of the
full Hamiltonian $H_{\rm sys}$.  We also explain how rotor symmetries
impose selection rules on the transition rates and how the resulting
equation is implemented in the numerical simulations. The system Hamiltonian is
invariant under reflection $\theta\mapsto -\theta$ and under a shift by
$\pi$, $\theta\mapsto\theta+\pi$. We model the environment as a set of independent bosonic baths indexed
by $\alpha$,
\begin{equation}
  H_{\rm B}
  = \sum_\alpha H_{\rm B}^{(\alpha)},\qquad
  H_{\rm B}^{(\alpha)} = \sum_k \hbar\omega_{\alpha k}
    b_{\alpha k}^\dagger b_{\alpha k},
\end{equation}
where $b_{\alpha k}$ annihilates a bath excitation of frequency
$\omega_{\alpha k}$ in channel  $\alpha \in \{ \cos\theta, \sin\theta,\cos2\theta,\sin2\theta,L_\theta\}$.  The system--bath coupling is
assumed bilinear,
\begin{equation}
  H_{\rm int}
  = \sum_\alpha S^{(\alpha)}\otimes B^{(\alpha)},
  \label{eq:Hint-app}
\end{equation}
with system operators $S^{(\alpha)}$ and bath operators
$B^{(\alpha)}$.  For the present problem the relevant system operators
are
\begin{equation}
  S^{(\cos\theta)} = \cos\theta,\qquad S^{(\sin \theta)} = \sin\theta,\qquad S^{(\cos2\theta)} = \cos2\theta,\qquad
  S^{(\sin 2\theta)} = \sin2\theta,\qquad
  S^{(L_\theta)}  = \hat L_\theta,
  \label{eq:Salpha-app}
\end{equation}
Each bath
operator is taken as a linear superposition of bosonic modes,
\begin{equation}
  B^{(\alpha)}
  = \sum_k g_{\alpha k}\bigl(b_{\alpha k} + b_{\alpha k}^\dagger\bigr),
  \label{eq:Balpha-app}
\end{equation}
 with coupling constants $g_{\alpha k}$. In the interaction picture with respect to $H_0 = H_{\rm sys} + H_{\rm B}$,
the interaction Hamiltonian is
\begin{equation}
  H_{\rm int}(t)
  = \sum_\alpha S^{(\alpha)}(t)\otimes B^{(\alpha)}(t),
\end{equation}
with the system and bath operators in the interaction picture,
\begin{align}
  S^{(\alpha)}(t)
  &= e^{+iH_{\rm sys}t/\hbar}
     S^{(\alpha)}
     e^{-iH_{\rm sys}t/\hbar},\\
  B^{(\alpha)}(t)
  &= e^{+iH_{\rm B}t/\hbar}
     B^{(\alpha)}
     e^{-iH_{\rm B}t/\hbar}.
\end{align}
The total density operator $\chi(t)$ obeys the von Neumann equation
\begin{equation}
  \frac{d}{dt}\chi(t)
  = -\frac{i}{\hbar}[H_{\rm int}(t),\chi(t)].
\end{equation}
Assuming factorized initial conditions
$\chi(0)=\rho(0)\otimes\rho_{\rm B}$ and tracing over the bath yields
the reduced density matrix $\rho(t)=\mathrm{Tr}_{\rm B}\chi(t)$.  Solving
the von Neumann equation to second order in $H_{\rm int}$ and inserting
back (Dyson expansion) leads to the Born equation
\begin{equation}
  \frac{d}{dt}\rho(t)
  = -\frac{1}{\hbar^2}\int_0^t d\tau\,
    \mathrm{Tr}_{\rm B}
    \left[
      H_{\rm int}(t),
      \bigl[H_{\rm int}(t-\tau),
            \rho(t-\tau)\otimes\rho_{\rm B}
      \bigr]
    \right].
  \label{eq:Born-app}
\end{equation}
In the Markov approximation, bath correlation times are short compared
with the system dynamics, so one may replace $\rho(t-\tau)\to\rho(t)$
and extend the upper integration limit to infinity
\begin{equation}
  \frac{d}{dt}\rho(t)
  = -\frac{1}{\hbar^2}\int_0^\infty d\tau\,
    \mathrm{Tr}_{\rm B}
    \left[
      H_{\rm int}(t),
      \bigl[H_{\rm int}(t-\tau),
            \rho(t)\otimes\rho_{\rm B}
      \bigr]
    \right].
  \label{eq:Born-Markov-app}
\end{equation}
Let $\ket{\psi_m}$ be the eigenstates of $H_{\rm sys}$,
\begin{equation}
  H_{\rm sys}\ket{\psi_m} = E_m\ket{\psi_m},\qquad
  m = 0,1,2,\dots,
  \label{eq:Hsys-eig-app}
\end{equation}
with Bohr frequencies $
  \omega_{mn} = ({E_m - E_n})/{\hbar}$. 
In this basis each system operator $S^{(\alpha)}$ can be expanded as
\begin{equation}
  S^{(\alpha)}
  = \sum_{m,n} S^{(\alpha)}_{mn}\ket{\psi_m}\bra{\psi_n},\qquad
  S^{(\alpha)}_{mn}
  = \bra{\psi_m}S^{(\alpha)}\ket{\psi_n}.
  \label{eq:S-mn-app}
\end{equation}
The interaction-picture operators then have the Fourier decomposition
\begin{align}
  S^{(\alpha)}(t)
  &= e^{+iH_{\rm sys}t/\hbar}
     S^{(\alpha)}
     e^{-iH_{\rm sys}t/\hbar} \nonumber= \sum_{m,n} S^{(\alpha)}_{mn}
     e^{+i(E_m - E_n)t/\hbar}
     \ket{\psi_m}\bra{\psi_n} \nonumber= \sum_\omega e^{-i\omega t} S^{(\alpha)}(\omega),
\end{align}
where we have grouped together all terms with the same Bohr frequency
$\omega=\omega_{mn}$ and defined the spectral components
\begin{equation}
  S^{(\alpha)}(\omega)
  = \sum_{\substack{m,n\\ E_m - E_n = \hbar\omega}}
      \ket{\psi_m}\bra{\psi_m}S^{(\alpha)}\ket{\psi_n}\bra{\psi_n}.
  \label{eq:Somega-app}
\end{equation}
By construction these satisfy the eigenoperator relation
\begin{equation}
  [H_{\rm sys},S^{(\alpha)}(\omega)]
  = -\hbar\omega\,S^{(\alpha)}(\omega).
\end{equation}
For a bath in thermal equilibrium, the bath operators enter Eq.~\eqref{eq:Born-Markov-app} only through
their equilibrium correlation functions
\begin{equation}
  C_{\alpha\theta}(\tau)
  = \mathrm{Tr}_{\rm B}\bigl[
      B^{(\alpha)}(\tau) B^{(\theta)}(0)\,\rho_{\rm B}
    \bigr],
\end{equation}
We assume that different channels
defined by the $S^{(\alpha)}$ are uncorrelated, 
$C_{\alpha\theta}(\tau)=\delta_{\alpha\theta}C_\alpha(\tau)$.  Using
Eq.~\eqref{eq:Balpha-app} and standard bosonic thermal averages one
obtains
\begin{equation}
  C_\alpha(\tau)
  = \int_0^\infty d\omega\,J_\alpha(\omega)
    \Bigl[
      \bigl(n_\alpha(\omega)+1\bigr)e^{-i\omega\tau}
      + n_\alpha(\omega)e^{+i\omega\tau}
    \Bigr],
  \label{eq:Calpha-app}
\end{equation}
Then the Born-Markov equation becomes
\begin{align}
  \frac{d}{dt}\rho(t)
  &= \frac{1}{\hbar^2}
     \sum_\alpha \int_0^\infty d\tau\,
     \biggl\{
       C_\alpha(\tau)\,
       \bigl[
         S^{(\alpha)}(t-\tau)\rho(t),S^{(\alpha)}(t)
       \bigr]
       + \mathrm{h.c.}
     \biggr\}. \label{eq:A-BM-C}
\end{align} where
\begin{equation}
  J_\alpha(\omega)
  = \sum_k |g_{\alpha k}|^2 \delta(\omega-\omega_{\alpha k})
\end{equation}
is the spectral density of bath $\alpha$, and
\begin{equation}
  n_\alpha(\omega)
  = \frac{1}{e^{\hbar\omega/k_{\rm B}T_\alpha} - 1}
\end{equation}
is the thermal occupation at temperature $T_\alpha$. We adopt the standard model spectral densities

\begin{align}
  J_\text{O}(\omega)   &= \eta_\text{O}\,\omega\,e^{-\omega/\omega_\text{O}},
                     &\text{(Ohmic)},\\
  J_\text{SO}(\omega)   &= \eta_\text{SO}\,\omega^3 e^{-\omega/\omega_\text{SO}},
                     &\text{super-Ohmic},\\
\end{align}
for $\omega>0$, with coupling strengths $\eta_\alpha$ and a high-frequency
cutoff $\omega_\alpha$.  For all decoherence mechanisms considered, the cut-offs $\omega_\alpha$ are largely irrelevant, as they are much higher than the considered tunneling rates and librational frequencies.  In the limit of thermal energies much smaller than the first librational energy $E_2$, the spectral densities are needed only at the tunnel-frequency, and thus boil down to a single decoherence rate that we estimate independently for the different decoherence mechanisms.  For larger temperatures, however, the scaling with $\omega$ of the spectral density can be used to estimate the jump rates $\gamma_\alpha$ defined below  for transitions to higher energy eigenstates. The one-sided Fourier transform of $C_\alpha(\tau)$ defines 
\begin{equation}
  \tilde\Gamma_\alpha(\omega)
  =\frac{1}{\hbar^2} \int_0^\infty d\tau\,e^{+i\omega\tau}C_\alpha(\tau).\label{gammaTil}
\end{equation}
We also define
\begin{equation}
\gamma_\alpha(\omega) = 2\,\mathrm{Re}\,\tilde\Gamma_\alpha(\omega),\qquad
\Lambda_\alpha(\omega) = \mathrm{Im}\,\tilde\Gamma_\alpha(\omega),
\end{equation}
such that
\begin{equation}
\tilde \Gamma_\alpha(\omega)
= \tfrac{1}{2}\gamma_\alpha(\omega) + i\,\Lambda_\alpha(\omega).\label{ReImGam}
\end{equation}
The real part $\gamma_\alpha(\omega)$ governs dissipative transitions as
\begin{equation}
  \gamma_\alpha(\omega)
  = 2\,\mathrm{Re}\,\tilde\Gamma_\alpha(\omega)
  = \frac{2\pi J_\alpha(|\omega|)}{\hbar^2}
    \begin{cases}
      n_\alpha(|\omega|)+1, & \omega>0,\\[4pt]
      n_\alpha(|\omega|),   & \omega<0,
    \end{cases}
  \label{eq:gamma-omega-app}
\end{equation}
while the imaginary part $\Lambda_\alpha(\omega)$ contributes to the
Lamb-shift Hamiltonian $H_{ LS}$. Inserting the spectral decompositions
$S^{(\alpha)}(t)=\sum_\omega e^{-i\omega t}S^{(\alpha)}(\omega)$ and the
correlation functions $C_\alpha(\tau)$ into
Eq.~\eqref{eq:Born-Markov-app}, performing the $\tau$ integral and
discarding rapidly oscillating terms with $\omega\neq\omega'$ (secular
approximation) yields
\begin{equation}
  \dot{\rho}
  = -\frac{i}{\hbar}[H_{\rm sys}+H_{\rm LS},\rho]
    + \sum_{\alpha,\omega}\gamma_\alpha(\omega)
      \left(
        S^{(\alpha)}(\omega)\rho S^{(\alpha)}(\omega)^\dagger
        - \frac{1}{2}
          \bigl\{
            S^{(\alpha)}(\omega)^\dagger S^{(\alpha)}(\omega),\rho
          \bigr\}
      \right),
  \label{eq:Lindblad-raw-app}
\end{equation}
with $\gamma_\alpha(\omega)$ given by
Eq.~\eqref{eq:gamma-omega-app}. We define
the jump operators
\begin{equation}
  L_{\alpha,\omega}
  = \sqrt{\gamma_\alpha(\omega)}\,S^{(\alpha)}(\omega),
  \label{eq:Lalpha-omega-app}
\end{equation}
and obtain the Gorini--Kossakowski--Lindblad--Sudarshan (GKLS) master
equation
\begin{equation}
  \dot{\rho}
  = -\frac{i}{\hbar}[H_{\rm sys}+H_{\rm LS},\rho]
    + \sum_{\alpha,\omega}
      \left(
        L_{\alpha,\omega}\rho L_{\alpha,\omega}^\dagger
        - \frac{1}{2}
          \bigl\{
            L_{\alpha,\omega}^\dagger L_{\alpha,\omega},\rho
          \bigr\}
      \right).
  \label{eq:Lindblad-final-app}
\end{equation}
This expression is valid for the full multi-level rotor and includes all decoherence channels.

\subsection{Numerical implementation in the angular-momentum basis}

For the numerical solution of Eq.~\eqref{eq:Lindblad-final-app} we
truncate the Hilbert space to angular-momentum states
$\ket{n}$, $n=-N,\dots,+N$, defined in Eq.~\eqref{eq:A1}.
In this basis the rotor
Hamiltonian $H_{\rm sys}$ is represented by a finite matrix $H_{nm}$ and
the operators in Eq.~\eqref{eq:Salpha-app} have simple analytic matrix
elements (e.g.~$\cos\theta$ couples $n$ to $n\pm1$ only).

Let $U$ denote the unitary matrix that diagonalizes $H_{\rm sys}$ in the
truncated basis,
\begin{equation}
 H_{\rm sys} = U E U^\dagger,
 \qquad
  E = \mathrm{diag}(E_0,E_1,\dots,E_M),
\end{equation}
so that the energy eigenstates are
\begin{equation}
  \ket{\psi_m} = \sum_n U_{nm}\ket{n}.
\end{equation}
The matrix elements of a system operator $S^{(\alpha)}$ in the energy
basis are then
\begin{equation}
  S^{(\alpha)}_{mn}
  = \bra{\psi_m}S^{(\alpha)}\ket{\psi_n}
  = \sum_{k,\ell} U_{mk}^* S^{(\alpha)}_{k\ell} U_{\ell n},
  \label{eq:S-transform-app}
\end{equation}
where $S^{(\alpha)}_{k\ell}$ are the matrix elements in the
$\{\ket{n}\}$ basis. For each channel $\alpha$ the jump operators
$L_{\alpha,\omega}$ in Eq.~\eqref{eq:Lalpha-omega-app} are constructed
as follows:
\begin{enumerate}
    \item Diagonalize $H_{\mathrm{sys}}$ to obtain eigenvalues $E_m$ and eigenvectors $|\psi_m\rangle$.
    \item Compute the matrix elements $S^{(\alpha)}_{mn}=\langle \psi_m|S^{(\alpha)}|\psi_n\rangle$.
    \item Construct the spectral components $S^{(\alpha)}(\omega)$ in the energy basis as in Eq.~\eqref{eq:Somega-app}. 
    In the numerical implementation, Bohr frequencies that agree within a chosen numerical tolerance are assigned to the same frequency.
    \item For each nonzero Bohr frequency $\omega\neq 0$, evaluate the transition rate, with the spectral density replaced with an effective coupling constant $\Gamma_{\alpha}$,
    \begin{equation}
    \gamma_{\alpha}(\omega)
    =
     \Gamma_{\alpha}
    \begin{cases}
    n_{\alpha}(|\omega|)+1, & \omega>0,\\[4pt]
    n_{\alpha}(|\omega|), & \omega<0.
    \end{cases}
    \label{eq:decratesgeneral}
    \end{equation}
    \item The zero-frequency contribution is treated separately, For an Ohmic bath
$
  J_\alpha(\omega) \sim \eta_\mathrm{O}\,\omega
$ 
at low frequency, while for small $\omega$ we have 
$
  n_\alpha(\omega)
  \sim ({k_{\rm B}T_\alpha})/({\hbar\omega}).
$ Then for
$\omega\to 0^+$,
\begin{equation}
  \gamma_\alpha(\omega)
  = \frac{2\pi J_\alpha(\omega)}{\hbar^2}\,[2n_\alpha(\omega)+1]
  \approx\frac{2\pi\eta_\mathrm{O}\omega}{\hbar^2}
  \cdot\frac{2k_{\rm B}T_\alpha}{\hbar\omega} + \mathcal{O}(\omega)
  =
  \frac{4\pi\eta_\mathrm{O} k_{\rm B}T_\alpha}{\hbar^3} + \mathcal{O}(\omega),
\end{equation}
and hence
\begin{equation}
  {
  \gamma_\alpha(0)
  \equiv \lim_{\omega\to 0}\gamma_\alpha(\omega)
  = \frac{4\pi\eta_\mathrm{O} k_{\rm B}T_\alpha}{\hbar^3}
  }
\end{equation}
for an Ohmic bath at finite temperature.
For a super-Ohmic bath, $J_\alpha(\omega)\sim\omega^3$, the product
$J_\alpha(\omega)n_\alpha(\omega)\sim\omega^2$ vanishes as
$\omega\to 0$, so $\gamma_\alpha(0)=0$ even at finite temperature. Numerically, $S^{(\alpha)}(0)$ is obtained from the diagonal part of $S^{(\alpha)}$ in the energy basis.
    
    \item The 
    jump operators passed to the master-equation solver are then
    \begin{equation}
    L_{\alpha,\omega}
    =
    \sqrt{\gamma_{\alpha}(\omega)}\,S^{(\alpha)}(\omega),
    \end{equation}
\end{enumerate}

The resulting set $\{L_{\alpha,\omega}\}$ is given as collapse
operators to the master-equation solver.  Because this construction is
performed entirely in the energy eigenbasis of $H_{\rm sys}$, it is
fully consistent with the Born-Markov-secular derivation and
automatically incorporates the selection rules associated with the
rotor symmetries.

\section{Eddy Currents}\label{app:eddy}

\subsection*{a. On the Particle}
The radius of possible particles in the setup is on the order of a few nm. We assume that the suspended particle is not a superconductor and that it is fully penetrated by the magnetic field in the trap. The magnetic field in the trap is
\begin{equation}
\mathbf{B}^{(m)}=
-\frac{\mu_0\zeta(3)}{4\pi L^3}
\begin{pmatrix}
2\mu_x\\
2\mu_y\\
3\mu_z
\end{pmatrix}
+\mathcal{O}(x^4),
\end{equation}
created by the surface currents on the superconducting plates. The particle's position is defined by the rotation of its magnetic moment
\begin{equation}
\boldsymbol{\mu}=\mu(\sin\theta,0,\cos\theta)^{\mathsf T}.
\end{equation}
Transforming this field into the particle's co-rotating frame yields the magnetic field actually seen by the particle
\begin{equation}
\mathbf{B}^{(p)}(\mu_x=\sin\theta,\mu_z=\cos\theta)=
-\frac{\mu_0\mu\zeta(3)}{8\pi L^3}
\left[
\begin{pmatrix}
0\\
0\\
5
\end{pmatrix}
+
\begin{pmatrix}
-\sin(2\theta)\\
0\\
\cos(2\theta)
\end{pmatrix}
\right]
=
\mathbf{B}^{(p)}_{\mathrm{dc}}+\mathbf{B}^{(p)}_{\mathrm{ac}}(2\theta).
\end{equation}
The magnetic polarizability of the particle material \(\alpha=\alpha'+i\alpha''\) yields the magnetic response of the particle with volume \(V\) to the applied magnetic field \(\mathbf{B}^{(p)}\), as
\begin{equation}
\boldsymbol{\mu}_p=\frac{1}{\mu_0}\alpha V \mathbf{B}^{(p)}.
\end{equation}
The static part of the magnetic field \(\mathbf{B}^{(p)}_{\mathrm{dc}}\) does not contribute to the average dissipated power. Thus, we only need to consider \(\mathbf{B}^{(p)}_{\mathrm{ac}}(2\theta)\). In the following, we model the trapped magnet as a particle rotating with constant angular frequency \(\omega\), so that \(\mathbf{B}^{(p)}_{\mathrm{ac}}(2\theta)=\mathbf{B}^{(p)}_{\mathrm{ac}}(2\omega t)\). Rewriting \(\mathbf{B}^{(p)}_{\mathrm{ac}}(2\omega t)\) in a complex representation, \(\mathbf{B}^{(p)}_{\mathrm{ac}}(2\omega t)=\tilde{\mathbf{B}}^{(p)}_{\mathrm{ac}}e^{-i2\omega t}\) with \(\tilde{\mathbf{B}}^{(p)}_{\mathrm{ac}}\propto(i,0,1)^{\mathsf T}\), we can calculate the average dissipated power \cite{landau1995},
\begin{equation}
P_\alpha
=
-\big\langle \boldsymbol{\mu}_p(t)\cdot \partial_t \mathbf{B}^{(p)}_{\mathrm{ac}}(t)\big\rangle_t
=
\frac{\mu_0\zeta(3)^2}{32\pi^2L^6}\,
\alpha''V\mu^2\omega ,
\end{equation}
with the time average denoted by \(\langle\cdots\rangle_t\) and
\begin{equation}
\alpha''=
\frac{1}{5}\left(\frac{R}{\delta}\right)^2,
\qquad
\delta=
\sqrt{\frac{2}{\mu_p\sigma\Omega}}.
\end{equation}
In this equation, \(\delta\) is the penetration depth of magnetic fields into normal conductors. It depends on the permeability of the particle \(\mu_p\approx\mu_0\), the angular frequency \(\Omega=2\omega\), and the conductivity \(\sigma\) of the particle. Using the relation between the dissipated power \(P_\alpha=\tau_\alpha\omega^2\) and the friction coefficient, we find the decay rate \(\Gamma^{\mathrm{eddy},\,p}=\tau_\alpha/I=P_\alpha/(I\omega^2)\), which is for low temperatures similar to the decoherence rate. For the considered geometry of the trap and particle properties (see Table \ref{tab:physical_correspondence}), we get
\begin{equation}
\Gamma^{\mathrm{eddy},\,p}=
\frac{2\pi\mu_0^2\zeta(3)^2}{135}\,
\frac{M_s^2\sigma R^{11}}{IL^6}
\sim 10^{-11}\,\mathrm{s}^{-1},
\end{equation}
with a conductivity of \(\sigma=0.667\,\mathrm{MSm}^{-1}\). Thus, eddy-current damping inside the particle is negligible on the timescales relevant to the tunneling dynamics.

\subsection*{b. On the Superconductor}
The London equations yield a finite penetration depth for magnetic fields into the superconductor, the London penetration depth $\lambda_\mathrm{L}$ \cite{Tinkham1996,London1935}. For type-I superconductors it is typically small. For example, tantalum has \(\lambda_L=150\,\mathrm{nm}\). Thus, we can model the currents in the superconductors as surface currents. They follow directly from the boundary condition for the magnetic field, as
\begin{equation}
\mathbf{J}=\hat{\mathbf{n}}\times\mathbf{B}/\mu_0=\mathbf{B}_\parallel/\mu_0
\end{equation}
with the surface normal vector \(\hat{\mathbf{n}}\) and the applied magnetic field \(\mathbf{B}\). In the setup the applied field is the field of the ferromagnetic particle and of the opposite wall. For example, considering the left superconductor,
\begin{equation}
\mathbf{B}(\mathbf{r})=\sum_{n}\mathbf{B}_n(\mathbf{r}),\qquad
\mathbf{B}_n(\mathbf{r})=
\frac{\mu_0}{4\pi}
\frac{
3\big(\boldsymbol{\mu}_n\cdot(\mathbf{r}-\mathbf{x}_n)\big)(\mathbf{r}-\mathbf{x}_n)-|\mathbf{r}-\mathbf{x}_n|^2\boldsymbol{\mu}_n
}{
|\mathbf{r}-\mathbf{x}_n|^5
},
\end{equation}
where \(n=0\) labels the particle and \(n\neq0\) are the image dipoles representing the induced currents for matching the boundary conditions. The positions of the dipoles are \(\mathbf{x}_n=(0,0,nL)^{\mathsf T}\) and their magnetic dipole moments are
\begin{equation}
\boldsymbol{\mu}_n=\mu(-ie^{i\omega t},0,(-1)^ne^{i\omega t})^{\mathsf T}.
\end{equation}
With the surface impedance of the superconductor \(Z_s=R_s+iX_s\) and \(\mathbf{E}=Z_s\mathbf{J}\), the average dissipated power from Ohm's law becomes
\begin{equation}
P_s=
\frac{1}{2\mu_0^2}
\Re\left\{
Z_s\int dA\,\langle \mathbf{B}_\parallel(t)\cdot\mathbf{B}_\parallel^*(t)\rangle_t
\right\}
=
\frac{I_0\mu^2}{32\pi^2L^4}R_s,
\qquad
I_0=
\frac{16\pi^2L^4}{\mu_0^2\mu^2}
\int_{\mathrm{Surface}} dA\,\langle \mathbf{B}_\parallel(t)\cdot\mathbf{B}_\parallel^*(t)\rangle_t.
\label{C7}
\end{equation}
The same mapping as for the case of the particle from the dissipated power to the decay rate holds for this case. We need to include an additional factor of \(2\) for accounting for the two plates. Thus, we find
\begin{equation}
\Gamma^{\mathrm{eddy,s}}=
\frac{2P_s}{I\omega^2}
=
\frac{I_0\mu^2}{16\pi^2\omega^2IL^4}R_s
=
\frac{I_0\mu_0^2}{18}\,
\frac{\sigma_n M_s^2R^6\lambda_L^3}{IL^4}\,
\frac{n_n}{n},
\end{equation}
with
\begin{equation}
\mu=M_sV=M_s\frac{4\pi R^3}{3},
\qquad
R_s=
\frac{1}{2}\frac{n_n}{n}\mu_0^2\sigma_n\omega^2\lambda_L^3.
\end{equation}
Here, \(I_0\) is the dimensionless geometric factor defined in Eq.~(\ref{C7}), $R_s$ is the surface resistance with conductivity $\sigma = \sigma_n\,n_n/n$, \(\sigma_n\) denotes the normal-state conductivity of the superconductor, and \(n_n/n\) denotes the normal-fluid (quasiparticle) fraction of the superconductor in a two-fluid description used for the conductivity $\sigma$ \cite{Tinkham1996,London1935,Gorter1934,turneaure1991}. It parametrizes the dissipative part of the electrodynamic response. In thermal equilibrium, \(n_n/n\) decreases rapidly for \(T\ll T_c\) (BCS: approximately exponentially in \(-\Delta/k_BT\)). For Ta with \(T_c\simeq 4\,\mathrm{K}\) at \(T=10^{-4}\,\mathrm{K}\), the equilibrium quasiparticle fraction is exponentially suppressed,
\begin{equation}
\frac{n_n}{n}\propto e^{-\Delta/(k_BT)},
\qquad
\Delta(0)\simeq1.76\,k_BT_c.
\end{equation}
Hence, the superconducting surface resistance, and thus the corresponding eddy-current damping rate, are negligible. 
For the parameters of Table \ref{tab:physical_correspondence}, $\lambda_L=150\,\mathrm{nm}$ and $\sigma_n \simeq 1 \, \mathrm{GSm}^{-1}$, one finds
\begin{equation}
\Gamma^{\mathrm{eddy,s}} \approx 1.07 \times 10^{5}\,\frac{n_n}{n}\ \mathrm{s^{-1}}.
\end{equation}

\section{Particle Scattering}\label{app:scattering}

The scattering of particles imprints information about the orientation of the particle in the environment if the particle is not perfectly rotationally symmetric. Following Carlesso et al.~\cite{carlesso_perturbative_2021}, we add a roughness of the particle characterized by the first spherical harmonics,
\begin{equation} \label{eq:PSpotential}
    r(\hat{\mathbf{r}}) = R \left[ 
    1+ \sum_i \frac{b_i}{R^3} Y_{1i}(\hat{\mathbf{r}})\right],
\end{equation}
where $b_i$ characterizes the roughness, defined by
\begin{equation}
    \frac{\Delta r}{R} = \frac{\sqrt{\int \mathrm{d}\Omega (r(\theta,\phi) - R)^2}}{2 \sqrt{\pi} R}\,. \label{eq:roughness}
\end{equation}
We translate the radial roughness  to a potential of the form
\begin{equation} 
    V(\mathbf{r}) = \sum_{l'',m''} d_{l'',m''}(|\mathbf{r}|) Y_{l'',m''}(\hat{\mathbf{r}}) = a \left[\kappa(r) Y_{00}(\hat{\mathbf{r}}) + \sum_{i=-1}^1 \frac{b_i}{r^3} Y_{1i}(\hat{\mathbf{r}})\right],\label{V}
\end{equation}
where $V(\mathbf{r}) \in \mathbb{R}$ yields $b_1 = -b_{-1}^\ast$. 
While in \cite{carlesso_perturbative_2021} a $1/r^3$ scaling of the potential was motivated by the study of  dipole-dipole interactions, we use \eqref{V} as a simple model that allows us to  
get an estimate of the order of magnitude of the decoherence rate, even if in reality the spatial dependence might be different.
The first term including $Y_{00}(\hat{\mathbf{r}})$ does not contribute as it is 
spherically 
symmetric. The first spherical harmonics $Y_{1i}(\hat{\mathbf{r}})$ characterize the derivation from the perfect spherical symmetry and will be considered as a model for the surface roughness. This yields according to Carlesso et al.~\cite{carlesso_perturbative_2021} the difference of the scattering amplitudes for a difference of angles $\Delta \theta = \theta - \theta'$,
\begin{equation}
    \Delta f^
    {\Delta\theta}(k \hat{\mathbf{k}}, k \hat{\mathbf{p}}) = f(k \hat{\mathbf{k}}, k \hat{\mathbf{p}},\theta) - f(k \hat{\mathbf{k}}, k \hat{\mathbf{p}},\theta') = -\frac{8 \pi m_\mathrm{gas}}{\hbar^2} \sum_{l,m} \sum_{l',m'} \sum_{l'',m''} 
    \mathcal{R}_{l,l',l'',m''}(k) \mathcal{G}_{l,m,l',m',l'',m''}(\Delta\theta)\, Y_{lm}^*(\hat{\mathbf{k}})\, Y_{l'm'}(\hat{\mathbf{p}})\,,
\end{equation}
where we have introduced the functions
\begin{equation}
    \begin{gathered}
        \mathcal{R}_{l,l',l'',m''}(k) = \int_0^\infty \mathrm{d}r \, r^2 j_l(kr) j_l'(kr) d_{l'',m''}(r),\\
        \mathcal{G}_{l,m,l',m',l'',m''}(\Delta\theta) = (-1)^{m'} i^{l' - l} \left(1 - e^{i \Delta\theta (m - m')}\right) \sqrt{\frac{(2l+1)(2l'+1)(2l''+1)}{4 \pi}} \begin{pmatrix} l & l' & l'' \\ m & -m' & m'' \end{pmatrix} \begin{pmatrix} l & l' & l'' \\ 0 & 0 & 0 \end{pmatrix},
    \end{gathered}
\end{equation}
with the Wigner 3j symbols and the spherical Bessel functions of the first kind $j_l$. Combining the above 
leads to 
the decoherence rate:
\begin{equation}
    n \int \mathrm{d}k \, v(k) \rho(k) \frac{\int \mathrm{d}\hat{\mathbf{k}} \int \mathrm{d}\hat{\mathbf{p}}}{8 \pi} {|\Delta f^
    {\Delta\theta}(k \hat{\mathbf{k}}, k \hat{\mathbf{p}})|}^2 = \int \mathrm{d}k \, v(k) \rho(k) \frac{96 m_\mathrm{gas}^2}{\pi^2 \hbar^4} n a^2 \Sigma \sin^2{\left(\frac{\Delta\theta}{2}\right)} (|b_1|^2 + |b_{-1}|^2), 
    \equiv \Lambda_R\sin^2{\left(\frac{\Delta\theta}{2}\right)}
\end{equation}
where $n = N / V$ is the particle number density, the constant $\Sigma$ is $\Sigma \approx \num{2.16}$, $v(k) = \hbar k / m_\mathrm{gas}$ is the velocity of the gas particles, and $\rho(k)$ is the momentum distribution. For a momentum distribution $\rho(k) = 4 \pi k^2 \mu(k)$, following from the Maxwell-Boltzmann distribution,
\begin{equation}
    \mu(k) = \left( \frac{\hbar^2}{2 \pi m_\mathrm{gas} k_\mathrm{B} T}\right)^\frac{3}{2} \exp{\left(-\frac{\hbar^2 k^2}{2 m_\mathrm{gas} k_\mathrm{B} T}\right)},
\end{equation}
we obtain the final result for the assumed potential as
\begin{eqnarray}
    \Lambda_\mathrm{R} = \frac{96\, m_\mathrm{gas}^2}{\pi^2 \hbar^4}\, n\, \Sigma\, a^2 \langle v \rangle  (|b_1|^2 + |b_{-1}|^2),\label{scattering_rate}
\end{eqnarray}
with $\langle v \rangle = \int \mathrm{d}k\, v(k) \rho(k) = \sqrt{8 k_\mathrm{B} T/(\pi m_\mathrm{gas})}$. For Helium ($m_\mathrm{gas} = \SI{4}{\atomicmassunit}$) at a temperature of $T = 3.2\,$mK, we obtain $\Lambda_\mathrm{R} \approx \SI{e85}{\J^{-2}\m^{-3}\s^{-1}} \cdot n \, a^2 \, |b_1|^2$. Thus, the roughness of the potential characterized by the contributions of the first spherical harmonics in Eq.~\eqref{eq:PSpotential} must fulfill $n \, a^2 \, |b_1|^2 \sim \SI{e-85}{\J^2\m^3}$ for decoherence rates $\Lambda_\mathrm{R} \sim \SI{1}{\per\s}$. 
Using Eq.~\eqref{eq:roughness} and assuming for simplicity $b_0=0$, we find
\begin{equation}
    \dfrac{\Delta r}{R}=\dfrac{|b_1|}{\sqrt{2\pi}R^3}
\end{equation}
such that we may directly calculate $|b_1|$. For example, $b_1 = 1.25\times \num{e-28}\,\text{m}^3$ yields a surface roughness of $\Delta r / R \approx \SI{5}{\percent}$ and the decoherence rate 
$\Lambda_\mathrm{R} \sim 
9.7\times\SI{e29}{\m^3\per\J^2\s} \cdot n \, a^2$.

To obtain an order-of-magnitude estimate for $a$, we match the strength of the isotropic part of the potential to an assumed elastic scattering cross section $\sigma$. Retaining only the $\ell=0$ term gives
\begin{equation}
V_{\mathrm{iso}}(\mathbf r)=a\,\kappa(r)\,Y_{00}(\hat{\mathbf r}), 
\qquad Y_{00}=\frac{1}{\sqrt{4\pi}}.
\end{equation}
Introduce the radial moment
\begin{equation}
I_\kappa \equiv \int_{0}^{\infty} dr\,r^2\,\kappa(r).
\end{equation}
Then the spatial integral of the isotropic potential is
\begin{equation}
\int d^3r\,V_{\mathrm{iso}}(\mathbf r)
= a\int_0^\infty dr\,r^2\kappa(r)\int d\Omega\,Y_{00}
= a\,I_\kappa\left(\int d\Omega\,\frac{1}{\sqrt{4\pi}}\right)
= a\,(2\sqrt{\pi})\,I_\kappa.
\end{equation}
In the first Born approximation the scattering amplitude is
\begin{equation}
f_B(\mathbf k'\!,\mathbf k)= -\frac{m_\mathrm{gas}}{2\pi\hbar^2}\int d^3r\;e^{-i\mathbf q\cdot\mathbf r}\,V(\mathbf r),
\qquad \mathbf q=\mathbf k'-\mathbf k,
\end{equation}
and in the low-momentum-transfer limit $q\to 0$ we approximate $e^{-i\mathbf q\cdot\mathbf r}\approx 1$, yielding
\begin{equation}
f_B(0)\approx -\frac{m_\mathrm{gas}}{2\pi\hbar^2}\int d^3r\,V_{\mathrm{iso}}(\mathbf r)
= -\frac{m_\mathrm{gas}}{2\pi\hbar^2}\,a\,(2\sqrt{\pi})\,I_\kappa
= -\frac{m_\mathrm{gas}\,a\,I_\kappa}{\hbar^2\sqrt{\pi}}.
\end{equation}
At low energy (s-wave dominated) the scattering amplitude is approximately angle-independent, 
$f_B(\theta)\approx -a_s$,where $a_s$ is the scattering length, and the total elastic cross section is
\begin{equation}
\sigma \approx 4\pi a_s^2.
\end{equation}
Identifying $a_s\equiv -f_B(0)$ gives
\begin{equation}
a_s=\frac{m_\mathrm{gas}\,a\,I_\kappa}{\hbar^2\sqrt{\pi}}
\quad\Rightarrow\quad
\sigma \approx 4\pi\left(\frac{m_\mathrm{gas}\,a\,I_\kappa}{\hbar^2\sqrt{\pi}}\right)^2
= \frac{4 m_\mathrm{gas}^2 a^2 I_\kappa^2}{\hbar^4}.
\end{equation}
Solving for $a$ yields the estimate
\begin{equation}
a \approx \frac{\hbar^2}{2 m_\mathrm{gas} \,I_\kappa}\,\sqrt{\sigma}. \label{aest}
\end{equation}

To obtain a concrete order-of-magnitude, we choose a minimal model for the isotropic profile,
$\kappa(r)=\Theta(R-r)$, so that
\begin{equation}
I_\kappa=\int_0^{R} r^2\,dr=\frac{R^3}{3}.
\end{equation}
We further take $\sigma$ to be of geometric/hard-sphere order, $\sigma\sim \pi R^2$ to $\sigma\sim 4\pi R^2$.
Inserting these choices into  
eq.\eqref{aest} 
gives
\begin{equation}
a \approx \frac{3\hbar^2}{2 m_\mathrm{gas} R^3}\sqrt{\sigma}
=
\frac{3\hbar^2}{2 m_\mathrm{gas} R^2}\times
\begin{cases}
\sqrt{\pi}, & \sigma=\pi R^2,\\[4pt]
2\sqrt{\pi}, & \sigma=4\pi R^2.
\end{cases}
\end{equation}
For $R=1\,\mathrm{nm}=10^{-9}\,\mathrm{m}$ and helium as the background gas ($m_\mathrm{gas}=4u=6.64\times 10^{-27}\,\mathrm{kg}$),
with $\hbar=1.0546\times 10^{-34}\,\mathrm{J\,s}$, this yields
\begin{equation}
a \approx
\begin{cases}
4.4 \times 10^{-24}\ \mathrm{J}, & \sigma\sim \pi R^2,\\[4pt]
8.8 \times 10^{-24}\ \mathrm{J}, & \sigma\sim 4\pi R^2.
\end{cases}
\end{equation}
Thus, at $R=1\,\mathrm{nm}$ a suitable order-of-magnitude estimate is
\begin{equation}
{
a \sim 10^{-23}\ \mathrm{J}.
}
\end{equation}
Inserting this into \ref{scattering_rate} yields 
\begin{equation}
    \Lambda_\mathrm{R} \sim 
  9.7  \times\SI{e-17}{\m^3\per\s} \cdot n,
\end{equation}
and taking the particle density for an ultra high vacuum to be around $n\sim 10^{11}\, \text{m}^{-3}$ we obtain a decoherence rate of
\begin{equation}
    \Lambda_\mathrm{R} \sim 
    9.7 \times10^{-6}\,\text{s}^{-1}.
\end{equation}

\section{Decoherence from phonons and seismic noise}\label{app:phonon}

Both acoustic phonons in the superconducting plates and classical vibrations of the trap center couple to the rotor orientation through the $L$-dependence of the trapping potential. Here we derive the resulting master equations and estimate the decoherence rates. The trapping potential~\cite{headley_magnetic_2025} for $\abs{z}\ll L$ has up to a constant the form $V=A(\theta)+B(\theta)\,z^2$, with
\begin{equation}
  A(\theta)=\frac{V_0}{2}(1+\cos 2\theta),\quad
  B(\theta)=\frac{K\,V_0}{L^2}(3+\cos 2\theta),
  \label{eq:potential_expansion}
\end{equation}
where $V_0=\mu_0\mu^2\zeta(3)/(8\pi L^3)$ and $K\equiv 93\,\zeta(5)/[4\,\zeta(3)]\simeq 20.06$.

We distinguish two types of phonons. The first type is related to fluctuations of the trap width $L$. They are denoted as acoustic phonons. The second type originates from a common fluctuation of both plates, corresponding to fluctuations of the trap center. It is called seismic noise in the following.\\

\subsection{Acoustic phonons}
\label{sec:acoustic_phonons}
We treat the superconductor as an isotropic elastic continuum with mass density $\rho_s$, longitudinal sound velocity $c_l$, and transverse sound velocity $c_t$. The relevant phonon wavelength at the tunneling frequency  and typical librational frequencies is much larger than the size of the trap. Thus, the particle only reacts to a change in the separation $L$ of the plates,
\begin{equation}
    L \rightarrow L + \delta L_{\mathrm{eff}}(t),
\end{equation}
where $\delta L_{\mathrm{eff}}(t)$ is an effective long-wavelength-average fluctuation seen by the trapped nanomagnet. We consider only the width fluctuation keeping the position of the particle fixed at $z_0 = 0$. Hence, the angular dependence of the potential is completely given by $A(\theta)$ in Eq.~\eqref{eq:potential_expansion}. Since $V_0\propto L^{-3}$, expanding the potential to first order in $\delta L_{\mathrm{eff}}$ gives
\begin{equation}
    H_{\mathrm{int}}^{(\mathrm{ac})} = \frac{\partial U(\theta)}{\partial L}\,\delta L_{\mathrm{eff}}(t)
        = -\frac{3V_0}{L}\cos^2\theta\,\delta L_{\mathrm{eff}}(t) 
        = -\frac{3V_0}{2L}\cos 2\theta\,\delta L_{\mathrm{eff}}(t) -\frac{3V_0}{2L}\delta L_{\mathrm{eff}}(t).
        \label{eq:APhononHint}
\end{equation}
The second term is proportional to the identity in the rotor Hilbert space and can be absorbed into the bath Hamiltonian. The system coupling agent in the interaction with acoustic phonons is therefore
\begin{equation}
    S_{\mathrm{ac}} = \cos 2\theta,
\end{equation}
with coupling strength
\begin{equation}
    g_{\mathrm{ac}} = \frac{3V_0}{2L}.
\label{eq:gac}
\end{equation}

Let the two opposite walls lie at $z=\pm L/2$, and denote their normal displacement fields by $u_{z}(-L/2,\mathbf r_\perp,t)$ and $u_{z}(+L/2,\mathbf r_\perp,t)$, where $\mathbf r_\perp = (x, y)$ is the in-plane coordinate of the walls. Because the magnetic trap samples a finite lateral patch, we define
\begin{equation}
    \delta L_{\mathrm{eff}}(t) = \int d^2 r_\perp\,\mathcal K(\mathbf r_\perp) \Bigl[ u_z(+L/2,\mathbf r_\perp,t) - u_z(-L/2,\mathbf r_\perp,t) \Bigr],
    \label{eq:dLeff}
\end{equation}
where $\mathcal K(\mathbf r_\perp) \in \mathbb{R}$ is a normalized weighting kernel satisfying $\int d^2r_\perp\,\mathcal K(\mathbf r_\perp)=1$. Its in-plane form factor is 
\begin{equation}
    F_\perp(\mathbf k_\perp) = \int d^2r_\perp\,\mathcal K(\mathbf r_\perp)e^{i\mathbf k_\perp\cdot\mathbf r_\perp}.
    \label{eq:formfactor}
\end{equation}
For each plate we expand the normal displacement field in acoustic modes considering one longitudinal in $z$ direction and two transversal modes in $x$ and $y$ direction,
\begin{equation}
    u_z(r_z,\mathbf r_\perp,t) = \sum_{\mathbf k,j} \sqrt{\frac{\hbar}{2\rho_s V\omega_{\mathbf k j}}}\, \mathbf e_{\mathbf k j,z} \left( b_{\mathbf k j}e^{-i\omega_{\mathbf k j} t} +  b^\dagger_{-\mathbf k j}e^{i\omega_{\mathbf k j} t} \right) e^{i(\mathbf k_\perp\cdot \mathbf r_\perp +k_z r_z)}.
    \label{eq:uzmode}
\end{equation}
In this equation, $V$ is the normalization volume, $j \in \{l,t_1,t_2\}$ labels the phonon branch, $\omega_{\mathbf k, j}$ is the phonon frequency, $b_{\mathbf k j}$ and $b^\dagger_{\mathbf k j}$ are bosonic annihilation and creation operators, $\mathbf e_{\mathbf k j,z}$ is the $z$ component of the polarization vector of the mode with $\mathbf e_{\mathbf k j,z}^\ast = \mathbf e_{-\mathbf k j,z}$. Inserting Eq.~\eqref{eq:uzmode} into Eq.~\eqref{eq:dLeff}, one obtains
\begin{equation}
    \delta L_{\mathrm{eff}}(t) = \sum_{\mathbf k,j} \ell_{\mathbf k j} \left( b_{\mathbf k j}e^{-i\omega_{\mathbf k j}t} + b^\dagger_{-\mathbf k j}e^{i\omega_{\mathbf k j}t} \right),
    \label{eq:dLeff_modes}
\end{equation}
with mode amplitudes
\begin{equation}
    \ell_{\mathbf k j} = 2i\sin\!\left(\frac{k_z L}{2}\right) F_\perp(\mathbf k_\perp) \mathbf e_{\mathbf k j,z} \sqrt{\frac{\hbar}{2\rho_s V\omega_{\mathbf k j}}}.
    \label{eq:ellkj}
\end{equation}
The derived mode expansion of the effective long-wavelength-averaged fluctuations $\delta L_\mathrm{eff}(t)$ yields the spectral density of the width fluctuations,
\begin{equation}
    J_\mathrm{L}(\omega) = \sum_{\mathbf l,j} \left|\ell_{\mathbf k j}\right|^2 \delta\!\left(\omega-\omega_{\mathbf k j}\right).
    \label{eq:JLdef}
\end{equation}
We can insert the coefficients $\ell_{\mathbf k j}$ into this equation and use the dispersion relation $\omega_{\mathbf k j} = c_j \abs{\mathbf{k}} = c_j k$ and $j=l,t_1,t_2$. The constant $c_j$ in the dispersion relation is the speed of sound for longitudinal modes $c_{l}$ and transversal modes $c_{t_1} = c_{t_2} = c_t$. This yields by replacing the $\sum_{\mathbf k}\to V\int d^3k/(2\pi)^3$ the general form of the spectral density,
\begin{equation}
    J_\mathrm{L}(\omega) = \frac{2\hbar}{\rho_s\omega} \sum_j \int\frac{d^3k}{(2\pi)^3} \sin^2\!\left(\frac{k_z L}{2}\right) \abs{F_\perp(\mathbf k_\perp)}^2 \abs{\mathbf e_{\mathbf k j,z}}^2 \delta(\omega-c_jk).
    \label{eq:JL_general}
\end{equation}
For the small trap size, the long-wavelength limit is relevant here, so the form factor may be set to unity $F_\perp(\mathbf k_\perp)\simeq 1$. The evaluation of the radial integral yields the spectral density
\begin{equation}
    J_\mathrm{L}(\omega) = \frac{\hbar\omega}{4 \pi^2 \rho_s} \left[c_l^{-3} \Phi_l\left(\frac{\omega L}{2 c_l}\right) + c_t^{-3} \Phi_t\left(\frac{\omega L}{2 c_t}\right) \right],
    \label{eq:JL_IlIt}
\end{equation}
where we used $\abs{\mathbf e_{\mathbf{k} l,z}}^2 = \cos^2(\theta)$ and $ \sin^2(\theta)= \sum_{j=t_1,t_2} \abs{\mathbf e_{\mathbf{k} j,z}}^2$. The functions $\Phi_l(a)$ and $\Phi_t(a)$ are the remaining angular integrals and can be evaluated exactly to
\begin{equation}
    \begin{aligned}
        \Phi_l(a) &= \frac{1}{\pi} \int d\Omega\,\cos^2(\theta)\sin^2(a \cos(\theta)) = \frac{2}{3} - \frac{\sin(2a)}{a} - \frac{\cos(2a)}{a^2} + \frac{\sin(2a)}{2 a^3},\\
        \Phi_t(a) &= \frac{1}{\pi} \int d\Omega\,\sin^2(\theta)\sin^2(a \, \cos(\theta)) = \frac{4}{3} + \frac{\cos(2a)}{a^2}-\frac{\sin(2a)}{2 a^3}.
    \end{aligned}
    \label{eq:PhiExact}
\end{equation}
In the long-wavelength limit we have $\omega L /c_j \ll 1$ for $j=j,t$. Thus, we can expand the exact solutions Eq.~\eqref{eq:PhiExact} for $a \ll 1$, which results in the final form of the spectral density
\begin{equation}
    J_\mathrm{L}(\omega) = \frac{\hbar L^2\omega^3}{60\pi^2\rho_s} \left(3c_l^{-5}+2c_t^{-5}\right) + \mathcal O\left(\omega^4 L^4 / c_j^4\right).
    \label{eq:JL_lw}
\end{equation}

Let us now derive from this result the spectral density following from the interaction Hamiltonian Eq.~\eqref{eq:APhononHint}. From Eq.~\eqref{eq:gac}, the microscopic coupling of the nanoparticle to the phonon mode $(\mathbf{k},j)$ is 
\begin{equation}
    g_{\mathbf k j} = \frac{3V_0}{2L}\,\ell_{\mathbf k j}.
    \label{eq:gmode}
\end{equation}
With this relation the acoustic spectral density associated with the system operator $S_{\mathrm{ac}}=\cos 2\theta$ is 
\begin{equation}
    J_{\mathrm{ac}}(\omega) = \sum_{\mathbf k, j} \left|g_{\mathbf k j}\right|^2 \delta\!\left(\omega-\omega_{\mathbf k j}\right) = \frac{9 V_0^2}{4 L^2} J_\mathrm{L}(\omega) = \frac{3 \hbar V_0^2 \omega^3}{80 \pi^2 \rho_s} \left(3c_l^{-5}+2c_t^{-5}\right),
    \label{eq:JacDef}
\end{equation}
with the definition of $J_\mathrm{L}(\omega)$, Eq.~\eqref{eq:JLdef}, and the final form of the width fluctuation spectral density Eq.~\eqref{eq:JL_lw}. Thus the acoustic bath is super-ohmic with $J_\mathrm{ac}(\omega) \propto \omega^3$. 
 We consider a bosonic bath at temperature $T$ resulting in the single-phonon emission- and absorption rates
 \begin{equation}
    \gamma_{\downarrow}(\omega) = \frac{2 \pi J_{\mathrm{ac}}(\omega)}{\hbar^2}\bigl(n(\omega)+1\bigr), \qquad \gamma_{\uparrow}(\omega) = \frac{2 \pi J_{\mathrm{ac}}(\omega)}{\hbar^2}n(\omega),\label{eq:gammaupdownx}
\end{equation}
As a benchmark we evaluate {these rates} at the tunneling frequency $\omega_\mathrm{T}$, where they become
\begin{equation}
    \gamma_{\downarrow}(\omega_\mathrm{T}) = \Gamma_{\cos{2\theta}}^{\mathrm{ac}}\bigl(n_\mathrm{T}+1\bigr),\qquad \gamma_{\uparrow}(\omega_\mathrm{T}) = \Gamma_{\cos{2\theta}}^{\mathrm{ac}} n_\mathrm{T},\qquad \mathrm{with} \qquad \Gamma_{\cos{2\theta}}^{\mathrm{ac}} = \frac{3V_0^2\omega_\mathrm{T}^3}{40 \pi \hbar\rho_s} \left( 3c_l^{-5}+2c_t^{-5} \right)
    \label{eq:gammadownfinal}
\end{equation}
and $n_\mathrm{T}\equiv n(\omega_\mathrm{T})$. For the benchmark parameters used in the paper, $R=1.0\times10^{-9}\,\mathrm{m}$, $L=8.4\times10^{-8}\,\mathrm{m}$, $V_0=1.779\times10^{-27}\,\mathrm{J}$, and $f_T=2.290\times10^5\,\mathrm{Hz}$. Further we consider a Tantalum superconductor for the trap with \cite{isbell1971} 
\begin{equation}
    \rho_s \simeq 16.65\times 10^3\,\mathrm{kg\,m^{-3}}, \qquad c_l \simeq 4.146\times10^3\,\mathrm{m\,s^{-1}}, \qquad c_t \simeq 2.032\times10^3\,\mathrm{m\,s^{-1}},
\end{equation}
one obtains
\begin{equation}
\frac{\omega_\mathrm{T} L}{c_t}\simeq 6\times10^{-5}, \qquad \frac{\omega_\mathrm{T} R}{c_t}\simeq 7.08\times10^{-7},
\end{equation}
so the long-wavelength approximation is very well satisfied. At the bath temperature $T=3.2\,\mathrm{mK}$, the phonon occupation at the tunneling frequency is $n_\mathrm{T} \simeq 290.7$. The microscopic acoustic rates are then
\begin{equation}
    \gamma_{\downarrow}(\omega_\mathrm{T})\simeq 2.249\times10^{-21}\,\mathrm{s}^{-1}, \qquad \gamma_{\uparrow}(\omega_\mathrm{T})\simeq 2.241\times10^{-21}\,\mathrm{s}^{-1},\qquad \Gamma_{\cos{2\theta}}^{\mathrm{ac}} = 7.714\times10^{-24} \,\mathrm{s}^{-1}
\label{eq:GammaNumeric}
\end{equation}
for the lowest tunneling doublet. However, the direct one-phonon transition between the two lowest states is forbidden to first order as $\langle\psi_1|\cos2\theta|\psi_2\rangle = 0$. Therefore Eq.~\eqref{eq:GammaNumeric} should be interpreted as the bath-induced channel strength at $\omega_\mathrm{T}$, while the actual decay of the protected tunneling doublet is even smaller and is controlled by coupling to higher excited states. In practice, the microscopic acoustic-phonon channel is negligibly weak for the benchmark geometry.

\subsection{Seismic noise}
Classical vibrations displace the trap center by $z \to z + \delta z(t)$. Expanding the $z^2$-dependent part of the trapping potential around the equilibrium position, $z = 0$, gives $B(\theta) \delta z^2$. Hence, the interaction Hamiltonian for the seismic noise is
\begin{equation}
H_{\mathrm{int}}^{(\mathrm{vib})}
=
\frac{K V_0}{L^2}(3 + \cos 2\theta)\,\delta z^2
=
\frac{K V_0}{L^2}\cos 2\theta\,\delta z^2
+
\frac{3 K V_0}{L^2}\,\delta z^2 .
\label{eq:Hint_vib_split}
\end{equation}
The second term in the r.h.s.~of \eqref{eq:Hint_vib_split} is proportional to the identity in the Hilbert space of the rotor and has no effect on the reduced rotor dynamics, except for an irrelevant shift of the energy. It is therefore dropped. The trap-center vibration channel is thus represented in the main-text master equation by the symmetry-preserving system operator
\begin{equation}
S^{(\mathrm{vib})}=\cos 2\theta .
\end{equation}
Since this noise is classical technical noise rather than a thermal bosonic bath, we absorb it directly into an effective Markovian rate $\Gamma^{\mathrm{vib}}_{\cos 2\theta}$. We define the (shifted) bath coupling agent $X_z(t)\equiv \delta z^2(t)-\langle\delta z^2\rangle_c$, where $\langle\ldots\rangle_c$ means the average over the classical noise ensemble, and 
the connected correlator of $X_z$ with a noise amplitude $A_z$, $\langle X_z(t)X_z(0)\rangle_c \simeq A_z\,\delta(t)$. This is the bath correlation function, so far still without the coupling constants $g_k\equiv KV_0/L^2$.  The coupling constants are independent of the "bath mode" $k$, meaning here independent of the frequency components contained in $X_z(t)$.  

We assume white noise for $\delta z(t)$.  The full bath correlation function, including the coupling constants, can then be taken as 
\begin{equation}
    C(\tau)=\frac{K^2 V_0^2}{ L^4}\,A_z \delta(\tau)\,. \label{Ctau}
\end{equation}
The amplitude $A_z$ is related to the noise spectral density $S_z$ and the bandwidth $\Delta f$ of the vibration isolation system by $A_z=2S_z^2 \Delta f$.  The factor 2 arises from from Isserlis' theorem, a.k.a.~Wick's theorem, $\langle 
\delta z^4\rangle_c=2\langle
\delta z^2\rangle_c$.
Insert \eqref{Ctau} into \eqref{gammaTil} to find 
\begin{equation}
\tilde{\Gamma}(\omega)=\frac{1}{\hbar^2}\left(\frac{K V_0 S_z}{L^2}\right)^2\Delta f\,,
\end{equation}
where the lower bound of the integral in \eqref{gammaTil} contributes with a factor 1/2.
Comparison with \eqref{ReImGam} gives
\begin{equation}
    \gamma(\omega)=\frac{2}{\hbar^2}\left(\frac{K V_0 S_z}{L^2}\right)^2\Delta f\,,
\end{equation}
independent of $\omega$.  Hence, the single-phonon emission and absorption rates due to seismic noise at the tunneling frequency have both the same value,
\begin{align}
    \gamma_\downarrow(\omega_\text{T}) & = \gamma(\omega_\text{T})\\
    \gamma_\uparrow(\omega_\text{T}) & = \gamma(-\omega_\text{T}).
\end{align}
If the seismic noise resulted from a true heatbath at temperature $T$, this would mean that the temperature is so high that the difference between $n_\text{T}(\omega_\text{T})$ and $n_\text{T}(\omega_\text{T}) +1$ is negligible, i.e.~$n_\text{T}(\omega_\text{T})\gg 1$.  This is indeed the limit of classical noise.  From $\gamma_\uparrow(\omega_\text{T}) =\Gamma_{\text{cos} 2\theta}^\text{vib}\,n_\text{T}(\omega_\text{T})$ one can, for a given temperature, extract the rate $\Gamma_{\text{cos} 2\theta}^\text{vib}$. 
For a temperature of $3.2$\,mK and the parameters in table \ref{tab:physical_correspondence} one finds $n_\text{T}\simeq 290.7$ and $\Gamma_{\text{cos} 2\theta}^\text{vib}\simeq 158/$\,s.  However, for comparison with the other dominant decoherence mechanism, namely particle scattering, one should compare directly $\gamma_\downarrow(\omega_\text{T}), \gamma_\uparrow(\omega_\text{T})$ with $\Lambda_\text{R}$, as these are the rates that enter directly the master equation.  For the same parameters, we find $\gamma_\downarrow(\omega_\text{T})\simeq \gamma_\uparrow(\omega_\text{T})\simeq $ 4.6 $\times 10^4/$\,s $\gg \Lambda_\text{R}$. 
Trap-center vibrations are therefore 
likely the dominant decoherence channel. 
Because of the quartic scaling, reducing $\sqrt{S_z}$ by one order of magnitude suppresses the rate by four orders of magnitude.
The steep quartic scaling means that vibration isolation is the decisive experimental control knob for this decoherence channel.
With excellent isolation
($\sqrt{S_z}\sim 10^{-13}\,$m$/\sqrt{\text{Hz}}$), the vibration
rate drops to $\gamma_{\uparrow,\downarrow}(\omega_\text{T}) \sim 4.6\times10^{-4}\,$s$^{-1}$, well below the tunneling
frequency, but remains the dominant decoherence mechanism.
 
\begin{table}[h]
\centering
\caption{Vibration-induced decoherence rate and coherence time versus
displacement spectral density ($\Delta f=10^3\,$Hz,
$V_0=1.779\times10^{-27}\,$J, $L=8.4\times10^{-8}\,$m).}
\label{tab:vib_scaling_app}
\begin{tabular}{lccc}
\hline\hline
Isolation scenario &
$\sqrt{S_z}\;[\mathrm{m/\!\sqrt{Hz}}]$ & $\gamma_{\uparrow,\downarrow}(\omega_\text{T})
\;[\mathrm{s^{-1}}]$ &
$\tau_{\mathrm{vib}}=1/\gamma_{\uparrow,\downarrow}\, [\mathrm{s}]$  \\
\hline 
\\
Standard cryostat         & $10^{-11}$ & $4.6\times10^{4}$  & $\sim\!2.17\times10^{-5}$     \\
Good passive isolation    & $10^{-12}$ & $4.6$  & $\sim\!2.17\times10^{-1}\,$  \\
Excellent isolation~\cite{deWit2019}
& $10^{-13}$ & $4.6\times 10^{-4}$  & $\sim\!2.17\times10^3\,$    \\
\hline\hline
\end{tabular}
\end{table}

\section{Readout-device back-action}\label{app:readout}
The coupling of the rotating magnetic dipole to a flux-sensitive readout device introduces decoherence through measurement back-action. We summarize the key results here; full derivations are given in Ref.~\cite{Mueller2025}.  We denote the readout-circuit inductance by $\mathcal{L}$ to distinguish it from the plate separation~$L$. When the pickup coil is read out by a SQUID, the relevant back-action is determined by the SQUID dynamics. Modeling the SQUID as a two-level system with level splitting $\Delta=\sqrt{\Delta_0^2+\varepsilon_0^2}$, where $\Delta_0$ is the tunneling amplitude of the SQUID and $\epsilon_0$ is a constant field contribution from the particle, persistent current $I_p$, and flux sensitivity $\tilde\zeta$~\cite{Mueller2025,romero-isart2012}, adiabatic elimination of the SQUID yields in the white-noise limit the master equation
\begin{equation}
  \dot\rho
  = -\frac{i}{\hbar}[H_{\rm sys},\rho]
    + \Gamma^{\mathrm{SQ}}_{\sin\theta}\;\mathcal{D}[\sin\theta]\,\rho\,,
  \label{eq:ME_SQUID}
\end{equation}
with
\begin{equation}
\Gamma^{\mathrm{SQ}}_{\sin\theta}
=
\frac{2\tilde{\zeta}^{\,2}\Phi_1^{\,2}\kappa_{\mathrm{SQ}}}{\Delta^{2}+\kappa_{\mathrm{SQ}}^{2}},
\qquad
\tilde{\zeta}=\frac{2\Phi_0 I_p}{\hbar},
\qquad
\frac{\kappa_{\mathrm{SQ}}}{2 \pi}
=
T_2^{-1}
+
\frac{2N_b+1}{2T_1},
\label{eq:SQUID_rate_maintext}
\end{equation}
where 
$N_b=[\exp(\hbar\Delta/k_B T_{\mathrm{SQ}})-1]^{-1}$ the thermal occupation at the SQUID operating temperature and $T_1$, $T_2$ the SQUID relaxation and dephasing times. Here,
\begin{equation}
    \Phi_1 = \int_\mathcal{A} \mathrm{d}\mathbf{A} \cdot \mathbf{B}(\mathbf{r}) \approx \mathcal{A} \, \hat{\mathbf{n}}\cdot\mathbf{B}(\mathbf{r}),
\end{equation}
denotes the geometry-dependent flux amplitude induced in the pickup coil by the
nanoparticle, and $\mathcal A$ denotes the pickup-coil area and $\hat{\mathbf{n}}$ its unit length normal vector. The approximation holds for small areas if the magnetic field $\mathbf{B}(\mathbf{r})$ is approximately constant over the coil area. In the experimentally relevant regime $\Delta \gg \kappa_{\mathrm{SQ}}$, Eq.~\eqref{eq:SQUID_rate_maintext} simplifies to
\begin{equation}
\Gamma^{\mathrm{SQ}}_{\sin\theta}
\simeq
\frac{2\tilde{\zeta}^{\,2}\Phi_1^{\,2}\kappa_{\mathrm{SQ}}}{\Delta^{2}}.
\label{eq:SQUID_rate_dispersive}
\end{equation}
Using representative parameters
$\Delta/(2\pi)=10\,\mathrm{GHz}$,
$I_p=0.5\,\mu\mathrm{A}$,
$T_1=T_2=1\,\mathrm{ms}$,
pickup-coil area $\mathcal A=10\,\mathrm{nm}\times10\,\mathrm{nm}$ and a distance of the pickup area of $L/2\sim\SI{1e-8}{\m}$ yielding $\Phi_1\sim10^{-22}\,$Wb, one finds
\begin{equation}
\Gamma^{\mathrm{SQ}}_{\sin\theta}\sim 10^{-35}\,\mathrm{s}^{-1}.
\label{eq:SQUID_rate_estimate}
\end{equation}
This is negligible compared to all other decoherence channels and to the tunneling frequency $f_T= 2.290\times10^5\,$Hz. The extreme smallness has two origins: the weak flux signal ($\Phi_1\sim10^{-22}\,$Wb) produced by the small magnetic moment of a $1\,$nm particle, and the large frequency mismatch between the SQUID ($\sim\!10\,$GHz) and the rotor ($\sim\!10^5\,$Hz), which provides a Lorentzian suppression of order $\kappa_{\mathrm{SQ}}/\Delta^2\sim 10^{-19}$.  The SQUID readout therefore introduces no measurable back-action on the tunneling dynamics. If the pickup geometry is chosen such that the coupling is proportional to $\cos\theta$ instead, the same estimate applies with the replacement $\sin\theta\rightarrow\cos\theta$. In that case the channel additionally benefits from the reflection-symmetry protection discussed in the main text.

\end{document}